\documentclass[%
 aip,
 amsmath,amssymb,
 reprint,%
]{revtex4-1}

\usepackage{graphicx}% Include figure files
\usepackage{dcolumn}% Align table columns on decimal point
\usepackage{bm}% bold math
\usepackage{lineno}% Enable numbering of text and display math
\usepackage[utf8]{inputenc}
\usepackage[T1]{fontenc}
\usepackage{mathptmx}
\usepackage{etoolbox}
\usepackage{multirow}
\usepackage{ulem}
\usepackage{threeparttable}%表格加注释
\usepackage{booktabs} %需要加载宏包{booktabs} 三线表宏包
\usepackage{color}
\usepackage[cmyk]{xcolor}

\makeatletter
\def\@email#1#2{%
 \endgroup
 \patchcmd{\titleblock@produce}
  {\frontmatter@RRAPformat}
  {\frontmatter@RRAPformat{\produce@RRAP{*#1\href{mailto:#2}{#2}}}\frontmatter@RRAPformat}
  {}{}
}%
\makeatother
\begin{document}

\preprint{AIP/123-QED}

\title[]{Performance assessment of the effective core potentials under the Fermionic neural network: first and second row elements}
% Force line breaks with \\
\author{Mengsa Wang}
\affiliation{Graduate School of China Academy of Engineering Physics, Beijing 100088, China}
\affiliation{National Laboratory of Computational Physics, Institute of Applied Physics and Computational Mathematics, Beijing 100088, China}
\email[Mengsa Wang's email: ]{wangmengsa22@gscaep.ac.cn}

\author{Yuzhi Zhou}
\affiliation{%
CAEP Software Center for High Performance Numerical Simulation, Beijing 100088, China
}%
\affiliation{Institute of Applied Physics and Computational Mathematics, Beijing 100088, China}
\email[Corresponding to: ]{zhou\_yuzhi@iapcm.ac.cn}

\author{Han Wang}%
\affiliation{National Laboratory of Computational Physics, Institute of Applied Physics and Computational Mathematics, Beijing 100088, China}
\affiliation{HEDPS, CAPT, College of Engineering, Peking University, Beijing, China}
\email[Han Wang's email: ]{wang\_han@iapcm.ac.cn}

\date{\today}% It is always \today, today,

\begin{abstract}
The rapid development of deep learning techniques has driven the emergence of a neural network-based variational Monte Carlo method (referred to as FermiNet), which has manifested high accuracy and strong predictive power in the electronic structure calculations of atoms, molecules as well as some periodic systems. Recently, the implementation of the effective core potential (ECP) scheme in it further facilitates more efficient calculations in practice. But there still lack comprehensive assessments on the ECP's performance under the FermiNet. In this work, we set sail to fill this gap by conducting extensive tests on the first two row elements regarding their atomic spectral and molecular properties. Our major finding is that in general the qualities of ECPs have been correctly reflected under the FermiNet. Two recently built ECP tables, namely ccECP and eCEPP, seem to prevail on the overall performance. Specifically, ccECP performs slightly better on the spectral precision and covers more elements, while eCEPP is more systematically built from both shape and energy consistency, and better treats the core polarisation. On the other hand, the high accuracy of the all-electron calculations is hindered by the absence of relativistic effects as well as the numerical instabilities in some heavier elements. Finally, with further in-depth discussions, we generate possible directions for developing and improving the FermiNet in the near future.
\end{abstract}

\maketitle

%\begin{quotation}
%The ``lead paragraph'' is encapsulated 
%\end{quotation}

\section{Introduction}	
Boosted by the exponential growth of computing power and the development of effective algorithms, electronic structure methods have been constantly evolved to help us better understand the behaviours of atoms, molecules and condensed matter \cite{Martin2016,Jensen2017}. More recently, utilizing the latest deep learning techniques, Pfau \textit{et. al.} proposed a Fermionic neural network-based variational Monte Carlo method \cite{ferminet} (dubbed as FermiNet) for the accurate simulations of many-body ground state. This new ansatz is more expressive and flexible compared to the traditional Slater-Jastrow form \cite{SlaterJastrow}, and has shown to be comparable to projector Monte Carlo methods in precision. Later on, the original framework has been further improved both in accuracy and efficiency \cite{spencer2020better,von2211self}, and modified to accommodate more scenarios including periodic systems, positronic chemical processes and excited states \cite{FermientElectronGas,pfau2023natural,cassella2024neural}. The above examples manifest the strong potential of this novel electronic structure method.

Despite these appealing features, FermiNet still has the same scaling as the traditional VMC, i.e., $Z^{5.5 \sim 6.5}$ with respect to the nuclear charge $Z$. Meanwhile, the effective core potential (ECP) scheme can alleviate this problem significantly by treating the nucleus and core electrons as an effective core, and has been a common practice in other electronic structure methods, for instance, Density Functional Theory (DFT) \cite{kohn1965self}, Hartree-Fock (HF) \cite{hf1930,slater1930} or some high level calculations like CCSD(T) \cite{paldus1999critical} and Quantum Monte Carlo (QMC) \cite{qmc}. More recently, Li \textit{et. al.} first incorporated the ECP method in the FermiNet \cite{li2022fermionic}. Soon afterwards, similar functionality has been implemented by the FermiNet team \cite{pfau2023natural}.

Even though Ref.~\cite{li2022fermionic} provides some ECP results on the 3d transition metal elements under the FermiNet, there still lack comprehensive assessments on the performance of mainstream ECPs, which would be of significant reference value both from theoretical and practical point of view. In this work, based on the following ECPs, namely Stuttgart (STU) ECP\cite{STUgroup1,STUgroup2,STUgroup3-7,STUgroup8}, Burkatzki-Filippi-Dolg (BFD) ECP\cite{BFD}, energy consistent correlated electron pseudopotential (eCEPP)\cite{ecEPP}, and correlation consistent ECP (ccECP)\cite{ccecp1,ccecp2,ccecp1s}, we have performed extensive tests on the atomic ionization potential (IP) and electron affinity (EA), as well as the dissociation energy (DE) of hydrides for the first two row elements(we adopt the convention that H and He make up the zeroth period). Meanwhile, we have also taken this opportunity to re-evaluate the ability of FermiNet in terms of the all-electron calculations.

From our spectral results, we found that the qualities of ECPs have been consistently reflected under the FermiNet, as the ECPs constructed from high-accuracy correlated calculations in general outperform those from single-particle methods both in accuracy and transferability. Among them, ccECP and eCEPP give the best performance on average. In particular, ccECP achieves slightly better precisions in many scenarios, and covers more elements. While eCEPP incorporates the shape and energy consistency in a systematic manner and better treats the core-valence correlation.  

In addition, the all-electron results, though still being very accurate for the light elements, start to deviate due to a lack of relativistic treatment. Such errors can overwhelm the chemical accuracy(1.594 $mE_{h}$) as early as we get to the second row. Besides, the all-electron calculations are more prone to the numerical instabilities, which could also lead to some extra errors based on our observation. 

Lastly, we provide further analysis on the cause of errors and numerical behaviors, and some broader perspectives on how to improve the FermiNet in the near future.

This paper is organized as follows. We briefly introduce the FermiNet framework and ECP scheme in Sec.~\ref{section:ecp}. Then major results are presented in Sec.~\ref{section:results}. More discussions and further perspectives are given in Sec.~\ref{section:discussion}. In the last section, we briefly summarise this work. Details on the hyperparameters, efficiency, precision and extra original data are presented in the Appendix.
 
\section{Method}
\label{section:ecp}
\subsection{Theoretical framework}

The many-body stationary Schrödinger equation under the Born-Oppenheimer approximation\cite{born1985quantentheorie} is the starting point here and for most electronic structure methods:
\begin{equation}
\label{eq:schroedinger}
\hat{H}\psi(\textbf{X})=E\psi(\textbf{X}),
\end{equation}
% \WH{we may start using the notation $\bm X$ appearing in (4) right in the eq. (1).}
where $\psi(\textbf{X})$ is the wavefunction of $N$ electrons, $\textbf{X}=(\textbf{x}_{1},\textbf{x}_{2},...,\textbf{x}_{N})$ denotes the state of all electrons and $\textbf{x}_{i}=\{\textbf{r}_{i}, \sigma_{i}\}$ represents the coordinates and spin of the $i$-th electron. The electronic Hamiltonian has the following form,

\begin{equation}
\label{eq:AEHamiltonian}
     \hat{H}=-\frac{1}{2}\sum_{i}\nabla_{i}^{2} + \sum_{i> j}\frac{1}{r_{ij}} - \sum_{iI}\frac{Z_{I}}{r_{iI}},
\end{equation}
where $r_{ij}$ denotes the distance between the $i$-th and $j$-th electron, $r_{iI}$ denotes the distance between the $i$-th electron and $I$-th nucleus, and $Z_{I}$ is the charge of the $I$-th nucleus. And throughout this paper we work in the atomic units.

Solving Eq.~\eqref{eq:schroedinger} for many-body systems still remains a central task for today's theoretical physicists and chemists. Successful electronic structure methods have been developed, improved and verified over time, such as DFT, CCSD(T) and QMC, though they all have their own strengths and weaknesses. 
Recently, the FermiNet team proposed a novel ansatz of the many-body wavefunction,  which consists Slater determinants composed of permutation-equivalent functions. In this work, we use the full determinant variant \cite{von2022self,fermi_git}, therefore, the general form of the full wavefunction can be illustrated as a linear combination of 
$k$ determinants: 
\begin{equation}
\begin{aligned}
\psi(\textbf{r}_{1}^{\uparrow},...,\textbf{r}_{n\downarrow}^{\downarrow})&=
\Sigma_{k} \omega_{k} (\mathrm{det}[\phi_{i}^{k \alpha}(\textbf{r}_{j}^{\alpha};\{\textbf{r}_{/j}^{\alpha}\};\{\textbf{r}^{\bar{\alpha}}\})]),
%&\times  \mathrm{det}[\phi_{i}^{k\downarrow}(\textbf{r}_{j}^{\downarrow};\{\textbf{r}_{/j}^{\downarrow}\};\{\textbf{r}_{\uparrow}\})]),
\end{aligned}
\end{equation}
where $\bar{\alpha}$ is $\downarrow$ if $\alpha$ is $\uparrow$ or vice versa, and $\phi_{i}^{k\alpha}(\textbf{r}_{j}^{\alpha};\{\textbf{r}_{/j}^{\alpha}\};\{\textbf{r}^{\bar{\alpha}}\})$ is the permutation-equivalent function implemented by a multilayer perceptron. The overall anti-symmetry is then guaranteed by the determinants.
In FermiNet, as an unsupervised neural network, its input consists solely of features of one and two electron positions.
The energy expectation value is  minimized with respect to the parameters of the network by the routines of VMC:
\begin{equation}
    \mathcal{L}(\theta) = \frac{\int d\textbf{X} \psi^*_{\theta}(\textbf{X})\hat{H}\psi_{\theta}(\textbf{X})}{\int d\textbf{X} \psi^*_{\theta}(\textbf{X})\psi_{\theta}(\textbf{X})} ,
\end{equation}
where $\theta$ denotes the parameters of the network. Along the lines, deep learning techniques play a key role in improving the accuracy and efficiency. As a result, FermiNet is competitive to the projector Monte Carlo methods, contrast to the common belief that VMC is less accurate \cite{ferminet,spencer2020better}.

\subsection{Effective core potentials}
The ECP scheme is based on the idea that the core electrons remain almost unchanged in most condensed matter scenarios. Hence, the bare nuclear potentials in Eq.~\eqref{eq:AEHamiltonian} can be replaced by the effective ones where core electrons and nucleus are packed all together:
\begin{equation}
\label{eq:ECPHamiltonian}
		\hat{H} = -\frac{1}{2}\sum_{i}^{N_{v}}\nabla_{i}^{2} + \sum_{i>j}^{N_{v}}\frac{1}{r_{ij}} + \sum_{i}^{N_{v}}V_{i}^{\text{ECP}}(r_{i}),
\end{equation}
where $N_{v}$ denotes the number of the valence electrons, implying that fewer electrons are handled. Other than that, the strong correlations and relativistic effects near the core are also avoided. 

All ECPs tested in this work take the semi-local form:
\begin{equation}
\label{eq:ECPForm}
	V_{i}^{\text{ECP}}(r_{i}) = V_{\text{loc}}(r_{i}) + \sum_{l=0}^{l_{\text{max}}} V_{l}(r_{i}) \sum_{m} |Y_{lm}\rangle \langle Y_{lm}|.
\end{equation}
The first one on the RHS is the local term, which reflects the Coulomb potential of a nucleus screened by the core electrons, plus the repulsion from core electrons seen in a longer distance. The second are the non-local terms, which project the pseudo wavefunction into spherical harmonics $Y_{lm}$ in the vinicity of the core regions (the $V_{l}$ is designed to be short-ranged). It can be loosely understand as the mimic of overlapping and interactions between the valence and core electrons near the core. For all ECPs used in this paper, $V_l$ is expressed as a linear combination of Gaussian functions:
\begin{equation}
	V_{l}(r) = \sum_{p=1}^{p_{\text{max}}}\beta_{l,p}e^{-\alpha_{l,p}r^{2}},
\label{eq:gaussform}
\end{equation}
where $\alpha_{l,p}$ and $\beta_{l,p}$ are the fitting parameters.

In general, these parameters are determined from two aspects: energy {consistency} and shape {consistency} \cite{BFD}. Briefly speaking, energy consistency demands the isospectrality (or near isospectrality) between the all-electron and pseudo Hamiltonians on some subspace of valence states. The targeted spectral properties may include neutral excitation energies, IPs and EAs, and with different spin configurations if possible \cite{dolg1996accuracy}. This usually underpins the overall precision or qualities of the ECPs. On the other hand, shape consistency requires the spacial characteristics(mostly the one-body density matrix) should be the same or close enough. This generally improves the transferability of the ECPs under different chemical environments \cite{hamann1979norm}. Also note that when constructing the ECPs in practice, it is necessary to take considerations from both sides, and do trade-offs between the accuracy, reproducibility and transferability during the optimization \cite{ecEPP,ccecp1,ccecp2,ccecp1s}.

In the rest of this section, we briefly introduce some details of the STU ECP\cite{STUgroup1,STUgroup2,STUgroup3-7,STUgroup8}, BFD ECP\cite{BFD}, eCEPP\cite{TN13,TN15,ecEPP}, and ccECP\cite{ccecp1,ccecp2,ccecp1s}. For all the ECPs tested in this paper, [He] core is applied to the first row of elements, and [Ne] core is applied to the second row of elements. 
% Note that ccECP also provide [He] core tables for the second row of elements. But to better compare between different ECPs, we use the [Ne] core ECPs second row of elements in this paper.
It is important to mention that solely the ccECP offers [He] core tables for the second-row elements. Therefore, to fairly compare different ECPs, we use the [Ne] core ECPs for the second row of elements in this paper.

%\begin{table*}[t]
%\centering
%\footnotesize
%\small
%\caption{\small {Summary of the main characters of the ECPs tested in this paper. "Methods" indicates the major electron structure methods for the all-electron calculations in the construction of the ECPs. "Non-singular" means whether or not the potential is singular at the core. "NA" denotes the absence of corresponding ECP table. All the ECPs are built on the relativistic calculations, but in a form without spin-orbit coupling.}}
%~\\
%\begin{tabular}{ c@{\hspace{0.5cm}} c@{\hspace{0.5cm}} c @{\hspace{0.5cm}}c@{\hspace{0.5cm}} c@{\hspace{0.5cm}} c @{\hspace{0.5cm}}c @{\hspace{0.5cm}}c@{\hspace{0.5cm}}c }
%			\hline
%			\hline
%			ECPs&Methods&\multicolumn{2}{c}{$l_{\text{max}}$}&Relativistic&Non-singular&Energy consistent&\multicolumn{2}{c}{Shape consistent} \\ 
%			\hline
%			&&Li-Ne&Na-Ar&&&&Li-Ne&Na-Ar\\
%			\hline
%			STU&HF&2&2&$\surd$ &$\times$& $\surd $&$\times$&$\times$\\   
%			BFD&HF&0&1&$\surd$ &$\surd$& $\surd $&$\times$&$\times$ \\
%			eCEPP&CCSD(T)&1&NA&$\surd$&$\surd$&$\surd$&$\surd$&NA \\
%			ccECP &CCSD(T)&0&1&$\surd$&$\surd$&$\surd$&$\times$&$\surd$ \\
			
%			\hline
%			\hline
%\end{tabular}
%\label{table:1}	
%\end{table*}

\subsubsection{Stuttgart ECP}
Stuttgart (STU) ECP is a relativistic energy-consistent ECP, offering both semi-empirical and $ab$ $initio$ varieties\cite{STUgroup1,STUgroup2,STUgroup3-7,STUgroup8}. We employed semi-empirical pseudopotentials for Li, Be, Na, and Mg. For the remaining elements in the first two rows, we employed $ab$ $initio$ pseudopotentials. Although STU ECP has included core polarisation potentials (CPPs) for some elements\cite{STUgroup1,STUgroup2}, considering that elements from groups 13 to 17 and Ne do not have corresponding CPP parameters\cite{STUgroup3-7,STUgroup8}, to ensure fairness, the CPP term is ignored in our calculations of STU ECP. Meanwhile, since eCEPP has the CPPs for all elements in its table and is built upon most modern calculations, we investigate the CPP effects from eCEPP. 

STU ECP is singular at the nucleus due to the bare Coulomb potential form of the local term:
\begin{equation}
V_{\text{loc}}(r) = -\frac{Z_{\text{eff}}}{r},
\end{equation}
where $Z_{\text{eff}} = Z - Z_{\text{core}}$ represents the effective charge. $V_l$ is composed of one Gaussian function, namely, $p_{\text{max}}=1$ for all tested elements except for F, Ne, and Cl, where $p_{\text{max}}=2$.
Parameters $\alpha_{l,p},\beta_{l,p}$ 's in Eq.~\eqref{eq:gaussform} are adjusted to experimental ionization energies of the single-valence electron ions in the semi-empirical pseudopotentials, while the parameters in  $ab$ $initio$ pseudopotentials are fitted to the total valence energy of a multitude of low-lying configurations of the neutral atoms or the singly charged ions obtained from Russell-Saunders(LS) coupled quasi-relativistic all-electron HF calculations. The maximum angular quantum number $l_{\text{max}}$ is 1 for Li and Be, and increases to 2 from B to Cl.

\subsubsection{BFD ECP}
BFD ECP is a non-singular energy-consistent ECP and is designed to reproduce the all-electron scalar-relativistic HF energies for multiple configurations, including low-lying excitations, the first and second cations and ground state anions \cite{BFD}. To eliminate the singularity at the origin, the local term takes the following form:
\begin{equation}
    \begin{split}
		V_{\text{loc}}(r)&=-\frac{Z_{\text{eff}}}{r}+\frac{\beta_{\text{loc},1}}{r}e^{-\alpha_{\text{loc},1}r^{2}}+\beta_{\text{loc},2}re^{-\alpha_{\text{loc},2}r^{2}}\\&+\sum_{k=3}^{k_{\text{max}}}\beta_{\text{loc},k}e^{-\alpha_{\text{loc},k}r^{2}}.
    \end{split}
\end{equation}
For the first row elements, $l_{\text{max}}=0, k_{\text{max}}=3, p_{\text{max}}=1$, and for the second row elements, $l_{\text{max}}=1, k_{\text{max}}=3, p_{\text{max}}=1$. Additionally, the first derivative of the potential is zero at the origin. And to ensure the potential is repulsive near the nucleus, it is further required to have a concave shape at the origin, i.e.,
\begin{equation}
\alpha_{\text{loc},3}\beta_{\text{loc},3}+\alpha_{l,p}\beta_{l,p}>0, \quad \forall l,p.
\end{equation}

\subsubsection{eCEPP}
eCEPP is a relativistic, non-singular, shape 
and energy consistent ECP \cite{ecEPP}, and is an extension of the correlated electron pseudopotentials (CEPP) by the same authors \cite{TN13,TN15}. The shape consistency is constructed based on the Multi-Conﬁguration Hartree-Fock results to match the one-electron density matrix and the long-range effects of core polarisation beyond certain radius. And the energy consistency is achieved by reproducing the excitation energies, IPs and EAs, from CCSD(T) calculations. The core polarisation effect has been effectively approximated by an additional term as:
\begin{equation}
    V^{\text{cpp}}=-\frac{1}{2}\left(1-e^{-(\frac{r}{r_{0}})^{2}}\right)^{4}\frac{\alpha}{r^{4}},
\end{equation}
where $r_{0}$ is the cut-off radius, and $\alpha$ is dipole polarizability of the effective core. Now eCEPP comprises three terms: the local term, the non-local terms and the core polarisation term. For constraints, eCEPP is not only bounded at the origin, but also requires that the first and second order derivatives are equal to zero. Finally, the eCEPP table is only available for elements from Li to F in the first two rows of the periodic table, though it also covers some transition metal elements. For the elements from Li to F, $l_{\text{max}}=1, k_{\text{max}}=6, p_{\text{max}}=6$. The local and non-local terms of eCEPP have the same form as the BFD ECP, though more Gaussian functions are needed to take care of the shape consistency and core polarisation effects. To make better comparison, we also investigated the case without the core polarisation, denoted as eCEPP*.

\subsubsection{ccECP}
ccECP is a scalar-relativistic, non-singular, energy-consistent ECP constructed mainly based on CCSD(T)\cite{ccecp1,ccecp2,ccecp1s}.
For B, C, N, O, their tables are parameterized mostly by minimizing the atomic spectral errors, and occasionally resorts to spacial information and diatomic binding curves \cite{ccecp1}. For the rest first row elements and all second row elements, the many-body energy consistencies plus the single-body norm-conserving conditions are employed \cite{ccecp1s,ccecp2}. Also the CPP has been intentionally left out for simplicity. The transferability has been checked and improved by the calculations of some other molecular properties. ccECP has the same semi-local form and constraints as the BFD ECP. For the first row elements, $l_{\text{max}}=0, k_{\text{max}}=3, p_{\text{max}}=1$, and for the second row elements, $l_{\text{max}}=1, k_{\text{max}}=3, p_{\text{max}}=2$.

\section{Results}
\label{section:results}
In order to better test various ECPs, we have implemented the ECP scheme on a recent version of FermiNet\footnote{This version of FermiNet is the one committed on May 31st, 2023 of the main branch. And we believe the subsequent modifications to the lastest FermiNet would not change any major conclusions of this manuscript.}. Compared to the FermiNet in the original paper \cite{ferminet}, this version has undergone significant improvements both in accuracy and efficiency. They are realized through simplifying and modularizing the neural network, maximizing GPU utilization, and reducing memory usage in the Jax framework. In the following, we refer to this version as the modified FermiNet, while the FermiNet used in the original paper \cite{ferminet,spencer2020better} as the reference FermiNet. The STU ECP, BFD ECP and ccECP are imported from the PYSCF package \cite{sun2018pyscf}. The parameters of eCEPP are from the supplementary material of Ref.~\cite{ecEPP} and eCEPP is manually implemented.

Except for few simple explanations, we reserve more detailed discussions to Sec.~\ref{section:discussion}. And here we focus on illustrating the numerical results.

For extra information, the details of the training process and the hyperparameters used are provided in the Appendix.~\ref{app:hyper}. Discussions on efficiency, envelope functions and precision are presented in Appendices~\ref{app:eff}, \ref{app:env} and \ref{app:prec}, respectively.

\subsection{Ground state energy}
In this section, we briefly illustrate the results on the ground state energies of atoms and CH$_4$ molecule from the all-electron calculations of the FermiNet.

In order to improve the stability of the training process, we pretrain the wavefunctions by $10,000$ iterations to match the HF results on the cc-pVDZ basis set, following by a $100,000$ iteration training process. 
For heavier atoms such as P, S, and Cl, we increase the iterations of training to 200,000. And we reduce the learning rate parameter $lr.delay$ from $10,000$ to $5,000$ to ensure the stability of the training for S.(More details on the parameter $lr.delay$ can be found in the Appendix~\ref{app:hyper}) For Cl, when $lr.delay$ is set to $5,000$, there is still numerical instability in the training process, so we further reduce $lr.delay$ from $10,000$ to $3,000$ during the training. Unless otherwise specified, all experiments use double-precision and adhere to the default hyperparameters outlined in Table~\ref{table:11} of Appendix~\ref{app:hyper}.

The calculated and experimental ground state energies are summarized in Table~\ref{table:2}. 
(We also provide the ground state energies from the ECPs calculations in Table~\ref{table:8}, and the all-electron and ECPs results for hydrides in Table~\ref{table:9} in the Appendix for completeness.)
% \WH{what kind of results are presented in the table?}
And the main observation is the deviations from the experiments are generally small, with an average value of 0.92 $mE_h$. However, we have also seen a significant growing of errors starting from Al to Cl. This should largely due to: (1) a lack of relativistic treatment in the all-electron model, and (2) numerical instabilities in the training, which will also show up again in other all-electron results. 

\begin{table}[ht]
    \centering
    \small
\caption{\small Ground state energies for the first two row atoms and CH$_4$. $\Delta E$ represents the differences between the results of FermiNet and experimental values.
% \WH{why called ``simplifed''?}  and the experimental values. 
MAD represents the mean absolute errors of $\Delta E$. The molecular structure of CH$_4$ is from Ref.~\cite{ferminet}. The number inside the curly braces represents how many tens of thousands of training iterations. If no citation is provided, the data in this work is obtained from calculations of the modified FermiNet.
}

    ~\\
    
\begin{tabular}{ c@{\hspace{0.4cm}} c @{\hspace{0.4cm}} c@{\hspace{0.4cm}}  c@{\hspace{0.4cm}}  }
	\hline
		\hline
		&FermiNet$(E_{h})$&Exact$(E_{h})$\cite{chakravorty1993ground}& $\Delta E_{1}(mE_{h})$\\ 
		\hline
		Li &-7.47786(2)\{10\}&-7.47806&0.20(2) \\
        Be&-14.66733(1)\{10\}&-14.66736 &0.03(1)  \\   
		B&-24.65378(2)\{10\}&-24.65391&0.13(2)  \\
		C&-37.84473(2)\{10\}&-37.845&0.27(2)   \\
		N&-54.58891(3)\{10\}&-54.5892& 0.29(3)   \\
		O&-75.06683(5)\{10\}&-75.0673&0.47(5) \\
		F&-99.73333(5)\{10\}&-99.7339 &0.57(5)\\ 
        Ne&-128.93736(6)\{10\}&-128.9376&0.24(6) \\
		Na &-162.25430(9)\{10\}&-162.2546&0.30(9)  \\
		Mg&-200.05255(10)\{10\} &-200.053&0.45(10)  \\  
		Al&-242.34466(12)\{10\}&-242.346&1.34(12)\\
		Si&-289.35676(17)\{10\} &-289.359 &2.24(17) \\
		P&-341.25716(16)\{20\}&-341.259 &1.84(16) \\
		S&-398.10740(17)\{20\}&-398.11&2.6(17) \\
		Cl&-460.14513(22)\{20\}&-460.148&2.87(22)   \\
		CH$_4$&-40.51410(4)\{10\}&-&- \\
		\hline
		MAD&&&0.92\\
		\hline
		\hline
    \end{tabular}
    \label{table:2}	
\end{table}

Besides, together with the results from reference FermiNet in Table \ref{table:refFerminet}, we see that the accuracy of the modified FermiNet  is somewhat improved under the same neural network configurations, but also with less iterations of training. Except for Li and Be atoms, the ground state energies from the modified FermiNet are also closer to the experimental values. While for Cl, two FermiNets give almost equivalent results, yet the training iterations in the modified FermiNet is much less than the reference FermiNet. These justify the improvements in the modified FermiNet.

\subsection{Ionization potential}
\begin{table*}[htp]    
%\centering
\caption{\small{The IPs results from the all-electron and different ECPs calculations. $\Delta E$ denotes the deviations from the experimental values. AE is the abbrevation for all-electron. $^\dagger$ denotes the inclusion of the approximated relativistic effects in the results of the all-electron calculations. * indicates that the CPP term has been removed from the eCEPP calculations. MAD$_{1}$ and MAD$_{2}$ represent the mean absolute deviations for the first and second row of elements, respectively. And MAD has all the elements considered. Bold figures indicate the closest results to the experimental IPs. Consistent notations have been employed throughout this paper.}}

~\\
    %\scriptsize
    \small
    \makebox[16.8cm][c]{
    \begin{tabular}{ c@{\hspace{0.3cm}}  c @{\hspace{0.3cm}} c @{\hspace{0.3cm}} c@{\hspace{0.3cm}}  c @{\hspace{0.3cm}} c @{\hspace{0.3cm}} c@{\hspace{0.3cm}}  c@{\hspace{0.3cm}}  c @{\hspace{0.3cm}} c @{\hspace{0.3cm}} c@{\hspace{0.3cm}}  c@{\hspace{0.3cm}}  c@{\hspace{0.3cm}}  c}
    \hline
    \hline
    &\multicolumn{7}{c}{Ionization potential$(mE_{h})$}&\multicolumn{6}{c}{$\Delta E(mE_{h})$}\\ 
    \cmidrule(lr){2-8}\cmidrule(lr){9-14}
    Element&AE&STU&BFD&eCEPP&eCEPP*&ccECP&Expt.\cite{koga1997atomic}&AE (AE$^\dagger$)&STU&BFD&eCEPP&eCEPP*&ccECP\\ 
    \hline
    Li&197.91 &-&- &-& -&-&198.14&-0.23 (-0.23)&-&-&-&-&-\\
    Be&\textbf{342.56} &344.70 &343.81&345.63&343.57 &343.98 &342.60&\textbf{-0.04} (-0.03)&2.1&1.21&3.03&0.97&1.38\\  
    B&\textbf{304.91} &302.28 &302.29&305.73&303.73 &304.15&304.95&\textbf{-0.04} (-0.07)&-2.67&-2.66&0.78&-1.22&-0.8\\
    C&\textbf{413.88}&411.07 &411.17& 414.63&412.85&414.02 &413.81&\textbf{0.07} (-0.14)&-2.74&-2.64&0.82&-0.96&0.21\\
    N&\textbf{534.67} &531.77 &532.20&535.14&533.48&534.99&534.12&\textbf{0.55} (-0.11)&-2.35&-1.92&1.02&-0.64&0.87\\
    O&500.25&498.61 &498.30&\textbf{500.46}&499.73&499.68&500.45&-0.2 (-0.2)&-1.84&-2.15&\textbf{0.01}&-0.72&-0.77\\
    F &640.74 &640.09 &639.02&640.80&640.17&\textbf{640.23}&640.28&0.46 (-0.22)&-0.19&-1.26&0.52&-0.11&\textbf{-0.05}\\
    Ne&794.44& 792.72&\textbf{792.67}&-&- 
    &793.56&792.48&1.96 (0.04)&0.24&\textbf{0.19}&-&-&1.08\\
    Na&188.30 &-&- &-& -&-&188.86&-0.56 (-0.35)&-&-&-&-&-\\
    Mg& \textbf{280.32}&278.50 &278.08&-&-&278.68&280.99&\textbf{-0.67} (-0.33)&-2.49&-2.91&-&-&-2.31\\
    Al&219.06  & 220.29&220.12&-&-&\textbf{220.07}&219.97&-0.91 (-0.96)&0.32&0.15&-&-&\textbf{0.1}\\
    Si& \textbf{299.38}& 300.09&300.21&-&-& 300.25&299.57&\textbf{-0.19} (-0.83)&0.52&0.64&-&-&0.68\\
    P& 387.12& \textbf{386.45}&387.17&-&-& 387.60&385.38&1.74 (-0.15)&\textbf{1.07}&1.79&-&-&2.22\\
    S&\textbf{380.07}&379.37 &377.46&-&-& 378.04&380.72&\textbf{-0.65} (-0.41)&-1.35&-3.26&-&-&-2.68\\
    Cl& 475.05&476.18 &475.09&-&-&\textbf{ 476.32}&476.55&-1.5 (-2.47)&-0.37&-1.46&-&-&\textbf{-0.23}\\
    \hline
    MAD$_{1}$&&&&&&&&\textbf{0.44} (0.13)&1.73&1.72&1.03&0.77&0.74\\
    MAD$_{2}$&&&&&&&&\textbf{0.89} (0.79)&1.02&1.70&-&-&1.37\\
    MAD&&&&&&&&\textbf{0.65} (0.44)&1.40&1.71&1.03&0.77&1.03\\
    \hline
    \hline
    \end{tabular}
}
    \label{table:3}
\end{table*}

\begin{figure*}[ht]
    \centering
    \includegraphics[scale=0.085]{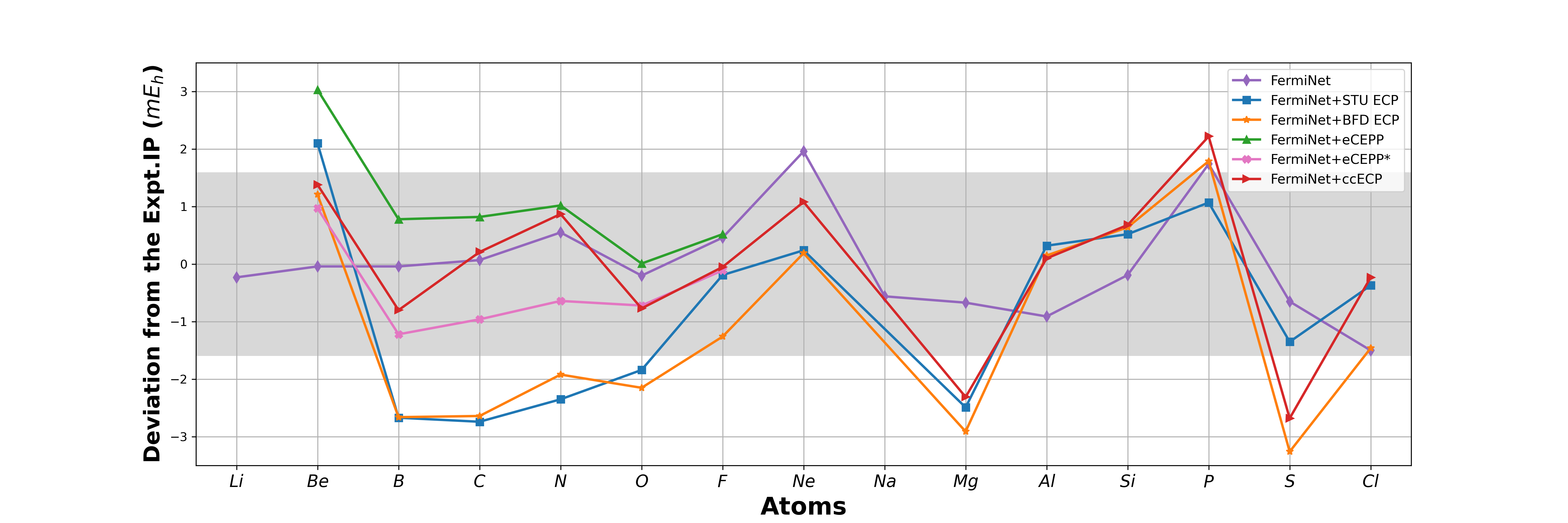}
    \caption{The plot of deviations of the IPs calculated by different ECPs and AE from the experimental IPs}
    \label{picture:1}
\end{figure*}	
Next we report the results on the IP, which is derived from the following formula:
\begin{equation}
    \text{IP}(\text{X}) = \text{E}(\text{X}^{+}) - \text{E}(\text{X}).
\end{equation}
Since the number of the valence electrons for Li$^{+}$ with [He] core and Na$^{+}$ with [Ne] core is zero, their ionization potentials are not considered in the current context. 
The experimental values \cite{koga1997atomic} serve as the reference data for both the all-electron and ECPs results.  As in the neutral atoms, to ensure a better convergence, we train $200,000$ iterations for the all-electron calculations of the P$^+$, S$^+$, and Cl$^+$. And we adjust $lr.delay$ from $10,000$ to $5,000$ for the S$^+$ and Cl$^+$ to improve the stability of the training process.

We summarize the IP results in Table~\ref{table:3} and Fig.~\ref{picture:1}. For all elements considered here, the all-electron results are on average closest to the experimental values, with a mean absolute error of 0.65 $mE_h$. Also note that for Ne, P and Cl, the errors of the all-electron become more significant, due to the relatively large relativistic effects \cite{klopper2010sub,koga1997atomic,maldonado2010quantum}, similar to what we have seen in the ground state energies.
Furthermore, as shown in Table~\ref{table:7} of Appendix~\ref{appendix1}, the all-electron IPs with relativity effects are even closer to the experimental IPs \footnote{Cl is an exception, whose error becomes even larger when incorporating the relativistic effects. This might be caused by the numerical issues as we have seen in Cl's training process.}, with a mean absolute error reducing from 0.65 to 0.44 $mE_h$. %\zyz{ccECP and eCEPP performs  best in the ECPs, with a mean absolute error of 0.7 $\sim$ 1.0 $mE_h$, though the latter only covers the first row.}

As for the ECPs results, we illustrate them row by row. For the first row, we observe that eCEPP and ccECP outperform the STU ECP and BFD ECP, and their absolute errors to experimental values are around 0.7 $\sim$ 1.0 $mE_h$. This is consistent with that eCEPP and ccECP are built from more accurate CCSD(T) calculations, and this character has been correctly reflected under the FermiNet. Meanwhile, the minor difference between the eCEPP and eCEPP* results indicates that the core polarisation term plays a less significant role in current situation.

\begin{table*}[t]
    \centering
    \caption{\small{The EAs calculated by different ECPs and the all-electron. $\dagger$ denotes the inclusion of the approximated relativistic effects in the results of the all-electron calculations. * indicates that the CPP term has been removed from the eCEPP calculations. $\Delta E$ denotes the deviations from the experimental EAs.}}
    %\scriptsize
    \small
    ~\\
    \begin{tabular}{ c @{\hspace{0.3cm}} c@{\hspace{0.3cm}}  c @{\hspace{0.3cm}} c @{\hspace{0.3cm}} c@{\hspace{0.3cm}}  c @{\hspace{0.3cm}} c@{\hspace{0.3cm}}  c@{\hspace{0.3cm}}  c@{\hspace{0.3cm}}  c@{\hspace{0.3cm}}  c @{\hspace{0.3cm}} c @{\hspace{0.3cm}} c@{\hspace{0.3cm}}  c }
		\hline
		\hline
		&\multicolumn{7}{c}{Eletron affinity$(mE_{h})$}&\multicolumn{6}{c}{$\Delta E(mE_{h})$}\\ 
		%\cline{2-9}
		\cmidrule(lr){2-8}\cmidrule(lr){9-14}
		Element&AE&STU&BFD&eCEPP&eCEPP*&ccECP&Expt.\cite{koga1997atomic}&AE (AE$^\dagger$)&STU&BFD&eCEPP&eCEPP*&ccECP\\ 
		\hline
		Li&21.49 &\textbf{22.70}&22.63&23.08&22.68&22.62&22.71&-1.22 (-1.23)&\textbf{-0.01}&-0.08&0.37&-0.03&-0.09\\
		%Be&- &-&-&-&- &- &-&-&-&-&-&-&-\\  
		B&9.59 &8.94 &9.56&\textbf{9.91}&9.73 &9.49&10.28&-0.69 (-0.65)&-1.34&-0.72&\textbf{-0.37}&-0.55&-0.79\\
		C&46.57&45.27 &45.48&\textbf{ 46.42}&46.08&46.71 &46.38&0.19 (-0.07)&-1.11&-0.90&\textbf{0.04}&-0.30&0.33\\
		%N&- &- &-&-&-&-&-&-&-&-&&-&-\\
		O&53.40&52.53 &53.07&54.09&\textbf{53.84}&54.27&53.70&-0.30 (-0.59)&-1.17&-0.63&0.39&\textbf{0.14}&0.57\\
		F &126.22 &124.15 &\textbf{124.87}&125.77&125.64&125.12&124.99&1.23 (0.26)&-0.84&\textbf{-0.12}&0.78&0.65&0.13\\
		%Ne&-& -&-&-&- &-&-&-&-&\textbf{0.19}&-&-&-\\
		Na&20.88&19.78&\textbf{20.17}&-& -&19.94&20.14&0.74 (0.74)&-0.36&\textbf{0.03}&-&-&-0.20\\
		%Mg&-&-&278.08&-&-&-&-&-&-&-2.91&-&-&-\\
		Al&14.88  & 16.67&16.57&-&-&\textbf{16.17}&15.91&-1.03 (-1.64)&0.76&0.66&-&-&\textbf{0.26}\\
		Si&52.97& \textbf{52.03}&52.04&-&-& 52.23&51.06&1.91 (1.12)&\textbf{0.97}&0.98&-&-&1.17\\
		P& 24.04& 25.23&25.65&-&-&\textbf{26.28}&27.44&-3.40 (-3.35)&-2.21&-1.79&-&-&\textbf{-1.16}\\
		S&74.36&76.11 &75.19&-&-&\textbf{76.33}&76.33&-1.97 (-2.67)&-0.22&-1.14&-&-&\textbf{0.00}\\
		Cl& 134.19&133.75 &\textbf{132.98}&-&-&134.27&132.76&1.43 (-0.56)&0.99&\textbf{0.22}&-&-&1.51\\
		\hline
        MAD$_{1}$&&&&&&&&0.73 (0.56)&0.89&0.49&0.39&\textbf{0.33}&0.38\\
        MAD$_{2}$&&&&&&&&1.75 (1.68)&0.92&0.80&-&-&\textbf{0.72}\\
		MAD&&&&&&&&1.28 (1.17)&0.91&0.66&0.39&0.33&\textbf{0.56}\\
		\hline
		\hline
    \end{tabular}
    \label{table:4}
\end{table*}

\begin{figure*}[ht]
    \centering
    \includegraphics[scale=0.085]{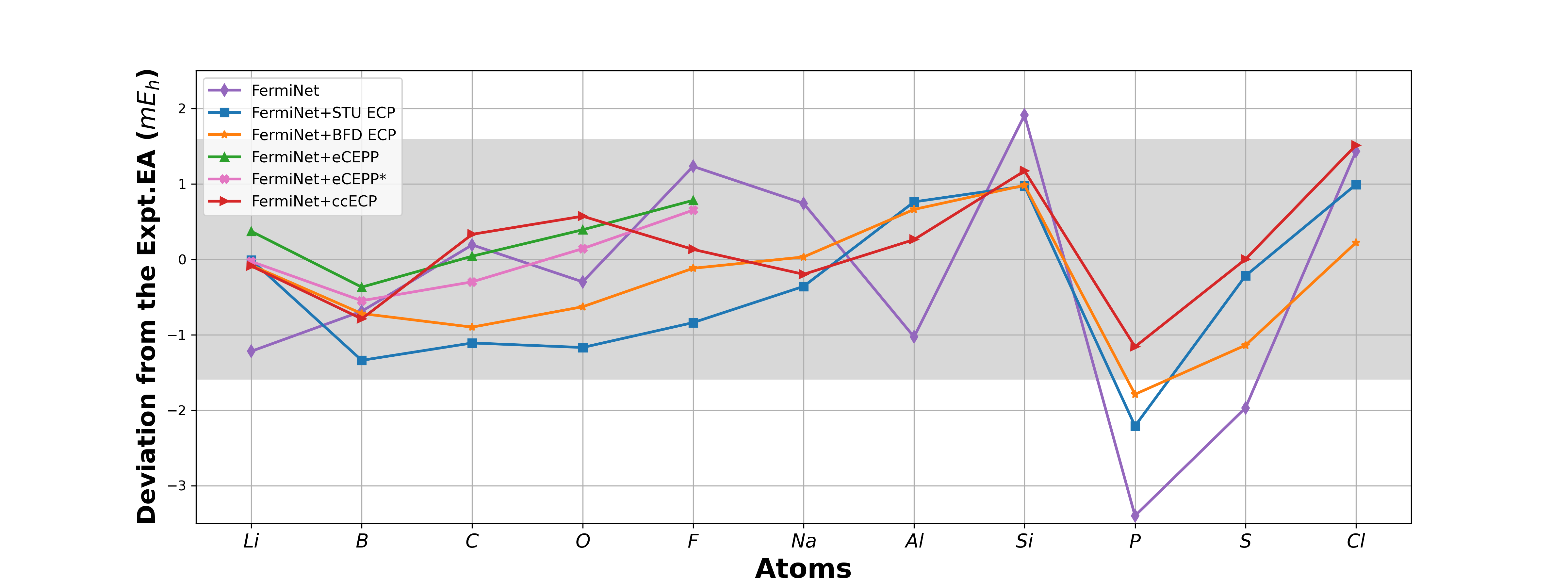}
    \caption{\small The plot of deviations of the EAs calculated by different ECPs and AE from the experimental EAs.}
    \label{picture:2}
\end{figure*}

For the second row, the advantage of ccECP disappears, and the results of the three ECPs are close to each other, as better seen in Fig.~\ref{picture:1}. For ccECP, we conjecture that it compromises the accuracy of spectral properties in order to ensure better transferability. On the other hand, we explain the relatively good performance of the STU ECP and BFD ECP in the following way. As observed in Ref.~\cite{ccecp2}, the ECP's correlation energies are almost rigid for particular states regardless of the construction. Consequently, the accuracy of the HF part of energies becomes more decisive, which underpins the accuracy of ECPs built from HF calculations.

\subsection{Electron affinity}
The prediction of EA is another important check on the performance of the ECPs and the all-electron calculations. The EAs of Be, N, Ne, and Mg are not considered due to the instability of their anions. The EA is calculated based on the following formula:
\begin{equation}
    \text{EA}(\text{X}) = \text{E}(\text{X}) - \text{E}(\text{X}^{-}).
\end{equation}
Similarly, we train the wavefunctions by $200,000$ iterations in the all-electron calculations of P$^-$, S$^-$ and Cl$^-$, and adjust $lr.delay$ to $5,000$ for S$^-$ and Cl$^-$. We use the experimental EAs \cite{koga1997atomic} as the reference data. All the EA results are summarized in Table~\ref{table:4} and Fig.~\ref{picture:2}.

It is somewhat surprising that the average deviation of the all-electron results is the largest, with a mean absolute error of 1.17 $mE_h$ and 1.28 $mE_h$, with and without relativistic effects.
Specifically, for Li, Al, Si, P and S, the deviations of the all-electron EAs with the relativistic effects (more detailed data can be found in Table \ref{table:7} of Appendix \ref{appendix1}) are all over 1 $mE_h$.  
Particularly, we find that convergence of Li$^-$ is slower compared to other small systems. We further train the wavefunctions of Li and its anion by additional $20,000$ iterations and the error of EA is reduced by 0.31 $mE_{h}$. And similar behaviors in Al, Si, P and S's EAs can be expected as we increase the training iterations or enlarge the neural networks.

\begin{table*}[htp]
    \centering
    %\footnotesize
    \small
    \caption{\small{The deviations of DEs of different ECPs compared to the all-electron results. $^\dagger$ denotes the inclusion of the approximated relativistic effects in the results of the all-electron calculations.* indicates that the CPP term has been removed from the eCEPP calculations. $\Delta E$ denotes the deviations from the all-electron results. }}
    ~\\	
    \begin{tabular}{ c@{\hspace{0.3cm}}  c@{\hspace{0.3cm}}  c@{\hspace{0.3cm}}  c@{\hspace{0.3cm}}  c @{\hspace{0.3cm}} c@{\hspace{0.3cm}}  c@{\hspace{0.3cm}}  c@{\hspace{0.3cm}}  c@{\hspace{0.3cm}}  c @{\hspace{0.3cm}} c@{\hspace{0.3cm}}  c }
		\hline
		\hline
		&\multicolumn{6}{c}{Dissociation energy$(mE_{h})$}&\multicolumn{5}{c}{$\Delta E(mE_{h})$}\\ 
		%\cline{2-9}
		\cmidrule(lr){2-7}\cmidrule(lr){8-12}
		Element&AE (AE$^\dagger$)&STU&BFD&eCEPP&eCEPP*&ccECP&STU&BFD&eCEPP&eCEPP*&ccECP\\ 
		\hline
		LiH &92.65&91.55 &91.88 &\textbf{93.37}&91.14& 91.36&-1.10&-0.77&\textbf{0.72}&-1.51&-1.29\\
		BeH$_2$& 236.00&\textbf{236.47} & 235.73&237.43&233.81&236.20 &0.47&-0.27&1.43&-2.19&\textbf{0.20}\\  
		BH$_3$&447.92& 446.01&\textbf{448.58}&450.07 &446.06& 447.18&-1.91&\textbf{0.66}&2.15&-1.86&-0.74\\
		CH$_4$& 669.37&665.89 &666.42&671.22&667.14& \textbf{669.91}&-3.48&-2.95&1.85&-2.23&\textbf{0.54}\\
		NH$_3$&474.46 & 470.64&471.31 &\textbf{474.87}&472.56&475.24&-3.82&-3.15&\textbf{0.41}&-1.90&0.78\\
		H$_2$O &371.04&368.75 &369.23&\textbf{371.48}&370.23 &370.05&-2.29&-1.81&\textbf{0.44}&-0.81&-0.99\\
		HF&225.69 (224.73\cite{visscher1996relativistic}) & 224.37&224.73&\textbf{225.25}&224.69&225.15 &-1.32&-0.96&\textbf{-0.44}&-1.00&-0.54\\
		NaH&71.81 &\textbf{71.78} &71.42 &-&-&72.37&\textbf{-0.03}&-0.39&-&-&0.56\\
		MgH& 50.41 &53.13 &53.95&-&-&\textbf{52.14}&2.72&3.54&-&-&\textbf{1.73}\\
		AlH &117.18 &118.08 &117.72&-&-&\textbf{117.28}&0.90&0.54&-&-&\textbf{0.10}\\
		SiH$_4$&514.44&517.98 &517.76&-&- &\textbf{516.46}&3.54&3.32 &-&-&\textbf{2.02}\\
		PH$_3$&380.84  &\textbf{381.02} &384.19&-&-&384.75&\textbf{0.18}&3.35&-&-&3.91\\
		H$_2$S& 287.59 (285.88\cite{peebles2002high})& 292.98&\textbf{291.44}&-&-&292.15 &5.39&\textbf{3.85}&-&-&4.56\\
		HCl&166.10 (164.19\cite{visscher1996relativistic}) &\textbf{170.86} &171.66&-&- &170.87&\textbf{4.76}&5.56&-&-&4.77\\
		\hline
		MAD$_{1}$&&&&&&&2.06&1.51&1.06&1.64&\textbf{0.73}\\
        MAD$_{2}$&&&&&&&\textbf{2.50}&2.94&-&-&2.52\\
        MAD&&&&&&&2.28&2.22&1.06&1.64&\textbf{1.62}\\
		\hline
		\hline
    \end{tabular}
    \label{table:5}
\end{table*}

\begin{figure*}[ht]
    \centering
    \includegraphics[scale=0.085]{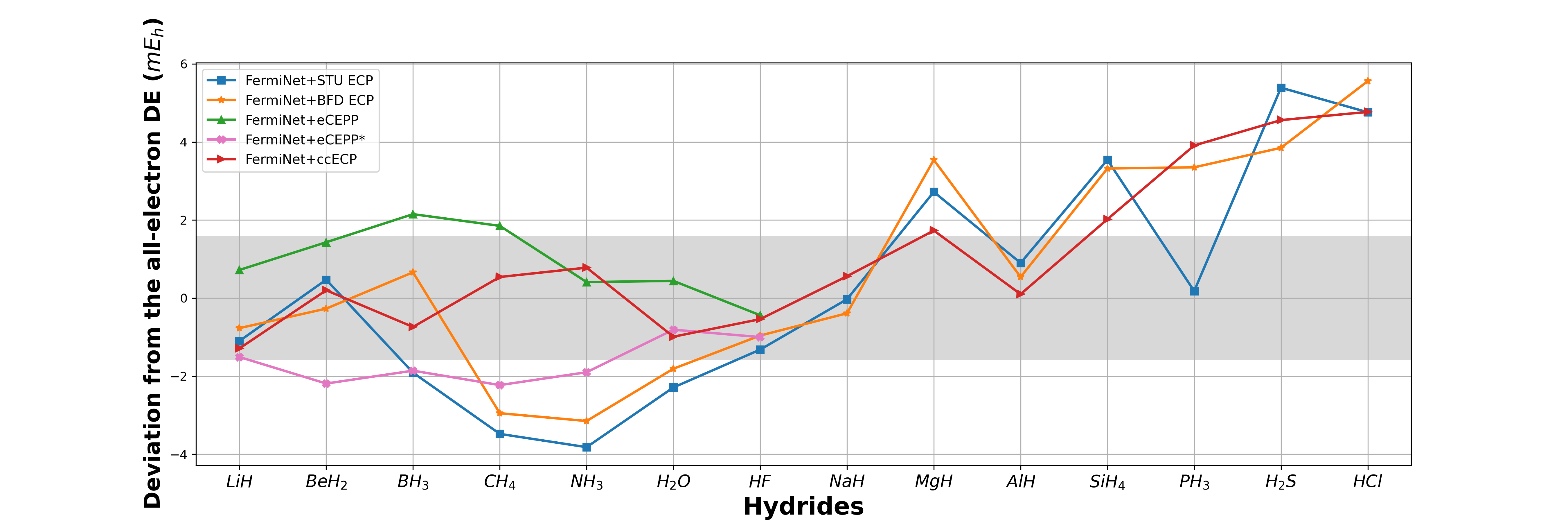}
    \caption{The plot of deviations of the DEs calculated by different ECPs from the all-electron DEs.}
    \label{picture:3}
\end{figure*}

In contrast, the ECPs give more accurate predictions on the EA. Among them, ccECP performs the best with a mean absolute error of 0.56 $mE_{h}$. While if only the first row elements is considered, it is the eCEPP* that has the smallest error of 0.33 $mE_{h}$. 
Those from STU ECP and BFD ECP are also within the chemical accuracy for all elements except for P. The slightly larger errors in STU ECP results should be atrributed to that they fit to the IPs and do not include EAs during the construction of the STU ECP. %Meanwhile, by comparing the results of BFD ECP and ccECP, it seems that the correlation energy does not have a significant impact on the accuracy of the EA. 
Meanwhile, the minor difference (0.06 $mE_{h}$) between eCEPP and eCEPP* indicates that the effect of core polarisation can be neglected also in this scenario.

We reserve more discussions on the cause of errors in the EA in Sec.~\ref{diss:ecp} and \ref{diss:ae} for the ECP and all-electron calculations, respectively. 
	
\subsection{Dissociation energy}
Lastly, we study the DE of hydrides from Li to Cl at the equilibrium states. The DE is calculated as follows:	
\begin{equation}
    \text{DE}(\text{X}_{n}\text{H}_{m}) = n\text{E}(\text{X}) + m\text{E}(\text{X}) - \text{E}(\text{X}_{n}\text{H}_{m}).
\end{equation}	
We set the ground state energy of the hydrogen atom to -0.5 $E_{h}$, and the molecule structures are from the CCCBDB experimental dataset \cite{CCCBDB} except for CH$_4$. We pretrain the all-electron wavefunctions of SiH$_4$, PH$_3$, H$_2$S and HCl by $20,000$ iterations. For hydrides with heavier elements, some literatures have given the estimates of the relativistic effects using different electronic structure methods \cite{tarczay2001anatomy,visscher1996relativistic,peebles2002high}. In Table~\ref{table:5}, we also give the DEs with the approximated relativistic effects for HF, H$_2$S and HCl. Moreover, we take the all-electron results as the reference to calculate the deviations, which are summarized in Table~\ref{table:5} and Fig.~\ref{picture:3} \footnote{The experimental results on DE can be found on Argonne Active Thermochemical Tables \cite{argonnethermo}, yet the experimental values contains many contributions from various reaction routes other than the simple one reflected in calculations. This will make the comparison less meaningful. Therefore, we refrain ourselves from comparing with the experimental DEs. Instead, we discuss the trends between the all-electron and ECPs results here.}.

The observations are as follows. Firstly, the poor consistency in the ECP results, as well as the large deviations to the all-electron results both indicate a variation of transferability. Yet, ccECP shows relatively good agreement with the all-electron up to Al. And if we assume the relativistic correction is small in these hydrides, ccECP seems to prevail in this contest, and eCEPP follows with slightly higher errors, about 0.3 $mE_h$ on average. This implies that either a systematic treatment of shape consistency as done in eCEPP \cite{TN13,TN15,ecEPP} or the incorporation of the spacial and molecular information in ccECP \cite{ccecp1,ccecp1s} do improve the transferability. In addition, for the second row hydrides, the results of STU ECP, BFD ECP and ccECP are close, and they start to deviate significantly from the all-electron results. This could, in part, be due to the relativistic effects in them, which are about 2 $mE_{h}$ in the cases of H$_2$S and HCl \cite{peebles2002high,visscher1996relativistic}. Some other part of the errors might come from numerical side, caused by more complex chemical environments in the hydrides. 

Moreover, unlike the situations with IP and EA, CPP is significant in improving the accuracy of DE. We observe a reduction of the mean absolute error by 0.58 $mE_{h}$ in eCEPP results when compared to eCEPP*. An more prominant example is LiH, which contains the most polarisable core among all hydrides considered here. (Since the relativistic effects is negligected for LiH, the all-electron results of LiH can serve as a good reference.) Including CPP reduces the error from -1.51 $mE_{h}$ to 0.72 $mE_{h}$, with a net effect of 2.2 $mE_h$. This aligns with the conclusion in Ref.~\cite{ecEPP}, which suggests that CPP significantly contributes to molecules containing Li. While ccECP, though being accurate in many other scenarios, gives the worst predictions here. %Therefore, in more complex molecular environments, the absence of the CPP term is one of the main sources of error for the other three types of ECPs.

\section{Discussion}
\label{section:discussion}
In this section, more in-depth discussions and analysis are given. Furthermore, we propose some future directions for improving the FermiNet.

\subsection{ECPs under FermiNet}
\label{diss:ecp}
Firstly, the overall assessment is that the qualities of the ECPs have been well preserved under the FermiNet. Among them, eCEPP and ccECP, which are most recently developed, prevail in most scenarios.
In particular, for IP and EA, the errors from these CCSD(T) based ECP tables are around the 1 $mE_h$. 
Meanwhile, the errors in the HF based ECPs (STU and BFD) are also not too far-off, about 0.1$\sim$1 $mE_h$ larger on average.
In the case of DE, the gaps between the HF and CCSD(T) based ECPs increase to 0.5$\sim$1.3~$mE_h$ on average. 
The DE results of hydrides also imply less consistency in the transferability. 
The incorporation of the core-valence correlation through CPP in eCEPP seems to have significantly improved the accuracy of DE, in contrast to the situations in IP and EA where CPP only presents a minor effect.

Other than that, the errors of EA are generally smaller than those of IP, which is also seen in the test results of ccECP papers \cite{ccecp1,ccecp2}. This again suggests that FermiNet has even maintained some subtle features of the ECPs, whether they have been constructed from HF or CCSD(T) calculations.  Meanwhile, the bias over the EA might due to the specific atomic valence states selected for the ECP construction, as the need to take care some highly positively charged ions might undermines the precision on the first IP. Also, a more flatness character of the ECPs makes the EA calculations much easier than those in the all-electron's, which will be soon discussed.

Subsequently, we check the valence-only correlation energies in the ground state, IP and EA, and compare with those from all-electron calculations using CCSD(T) \cite{ranasinghe2015core} (detail numbers shown in Table~\ref{table:valcorr} of Appendix~\ref{appendix2}). As anticipated and pointed out by Dolg \cite{valen-only}, for the ground state energies of atoms, ECP results generally overestimate the valence correlation energies in magnitude (though for N ccecp and eCEPP slighltly underestimate). The reason is that pseudo wavefunctions are more flat and with less nodes, which increases the chance for the like spins to overlap. However, we unexpectedly found that ccECP does no better job than eCEPP, or even BFD in some scenarios, although ccECP has more lean on the correlation energy consistency during the construction.

As for the spacial information, since there are difficulties to extract the corresponding pseudo and all-electron orbitals from FermiNet's MLP structure, we check the total densities instead. Only negligible differences in densities have been observed between different ECPs. Also the ECP densities converge to the all-electron's beyond certain core region. Representative density distributions of Carbon and LiH from ECP and all-electron calculations have been shown in Figs.~\ref{picture:density_C} and \ref{picture:density_LiH} in the appendix. Finally, we stress that this is only a glimpse rather than a strict assessment on the shape consistency.

%\wms{For ground state, the valence correlation energies from ccECP and eCEPP*, constructed using correlated methods, are closer to the all-electron valence correlation energies calculated using CCSD(T)\cite{ranasinghe2015core} compared to the results from STU ECP and BFD ECP (as shown in Table~\ref{table:valcorr} of Appendix~\ref{appendix2}).} 

%\wms{ Among them, eCEPP and ccECP, which are most recently developed, prevail in most scenarios. Based on the findings of this paper, CPP can be almost neglected in the calculations of IP and EA for the elements we tested. However, for the dissociation energies of hydrides, incorporating core-valence correlation through core polarisation term leads to more accurate descriptions.}

Lastly, we make some outlooks. Note that the ECPs tested here are all expressed as the combination of Gaussian functions, which facilitates the electronic structure calculations in the similar basis sets. On the positive side, this form poses no extra computation overhead under the FermiNet's framework, as suggested by our investigations on the efficiency (more in Appendix.~\ref{app:eff}). However, for heavier elements, one needs more and more Gaussians to express the more complex local part, as well as more channels in the non-local projections. This leads to a much harder optimization problem with very complicated constraint conditions. Therefore, it remains an open question whether we can take advantage of the strong expressibility of the neural network and design ECP specifically for FermiNet? This might be an interesting starting point for future studies.

\subsection{The all-electron results}
\label{diss:ae}
First and foremost, the lack of relativistic effects is a major issue in the all-electron calculations, and we will focus on it in Sec.~\ref{diss:rel}. As for now, we discuss on some other perspectives. Contrary to the ECPs results, the IPs from the all-electron calculations are in better agreement with the experimental data. While the EAs are generally worse, even with the relativistic effects being considered. This observation can be explained as follows. In anions, the added electron is loosely bounded and has a very different form of the wavefunction compared to the other strongly bounded electrons. Describing them is then a much difficult problem, which is known for the community for some time \cite{EAisDifficult}, and FermiNet is not an exception as well. 
% \WH{but we see the ferminet's IPs are more accurate than its EAs}
{However, in cations, all electrons are strongly bounded in nature, thus make the all-electron results of IP almost free from the above issue.}

Furthermore, to our relief, we found that even for some recent VMC/DMC calculations \cite{qmcbench}, the errors of EA are similar or sometimes even larger compared to our results. Meanwhile, CCSD(T) calculations, when extrapolating to the complete basis set (CBS) limit, give very close results to the experimental data \cite{ranasinghe2013ccsd,ccsdbench}. Yet for some intermediate basis sets, they still exhibit similar errors. This is consistent with the belief that FermiNet is equivalent or superior to VMC/DMC, but slightly lesser than the golden standard CCSD(T) \cite{ferminet}. This also implies that a better starting point from pretraining or a larger-scale neural network would most likely to improve the FermiNet's results.

On the other hand, as mentioned previously and also in Appendix~\ref{app:prec}, we have seen that the all-electron calculations are more prone to the numerical instabilities. In particular, some sudden rises in the energies can be observed during the training, and lead to some persistent errors at last. This can be better seen in the Cl's case of Fig.~\ref{picture:7}. Based on our further tests, they can be improved by more conservative learning strategies or more iterations. It is apparent that this phenomenon is related to the relaxation of hyperparameters into some meta-stable regions during the optimization, yet more detailed studies on the numerical behaviors of the FermiNet are needed to better understand the cause and find ways to avoid it.

\subsection{Relativity: Yes, but how?}
\label{diss:rel}
Last but not the least, the relativistic effects show up as early as we move into the second row, and render the all-electron results deviate from the experimental values by few $mE_h$. 
As discussed in appendix~\ref{appendix1}, current relativistic corrections are combinations of literature results, analytical and perturbative calculations at CCSD(T) and HF levels, which is apparently unsystematic. X2C corrections at HF level are also available along side. Still this scheme is unsatisfactory for high accuracy calculations. Therefore, more systematic treatments of the relativistic Hamiltonians under the FermiNet should undoubtedly be a critical issue. 

% Therefore, systematically implementing the relativistic model into the FermiNet should undoubtedly be the key improvement in future.

Then we have to choose the technical roadmap. A full, rigorous treatment of many-body system would require the Quantum Electrodynamics (QED) framework, which properly describes all processes between electrons and positrons with
photons and the external potentials. Yet even for atoms with few electrons, calculations taking account the low-order QED effects are very difficult, if not impossible \cite{relativChemBook}. From the other limit, we could start from the Schrödinger side and put perturbative terms arising from expanding the Dirac equation in the low energy limit, for instance, the Zeroth-Order Regular Approximation (ZORA). Though sharing a similar form with Schrödinger equation, such approximations would soon become ineffective as we move to heavier atoms (remember our goal is sub chemical accuracy!). Therefore, we envision that a practical solution should lie somewhere in between. As done in many other relativistic quantum chemistry calculations, the Dirac-Coulomb Hamiltonian seems to be a good starting point, and its relatively simple form also fits to VMC calculations. Alternatively, we could resort to some simplified 2-component Hamiltonians like X2C\cite{X2C} and $n$-order Douglas–Kroll-Hess(DKH$n$) Hamiltonians, which are computationally more tractable. However, we have to point out that either treatments fundamentally change the Hamiltonian and wavefunction. As a result, solving them calls for a substantial reformulation and reconstruction of the current framework, and is generally much harder than solving the Schrödinger equation.

Furthermore, if we do not restrict ourselves to the all-electron calculations, relativistic ECPs can be thought as a good choice to incorporate the relativistic effects. The benefits of doing so come in twofold. First, since the relativistic calculations are even more demanding than the non-relativistic ones, the gain in the computational cost reductions is even more significant from these ECPs. Second, because the relativistic ECPs generally take the form of spin-orbit coupling \cite{ccecp4}, it is then only necessary to modify current FermiNet to accommodate for variable spins. The extension in the traditional VMC calculations has been discussed in Ref. ~\cite{vmcspin}, which should cover most technique details, though minor modifications to FermiNet ansatz should also be considered.

Finally, we believe these two routes are of equal significance and are complementary to each other in the practical applications. Thus both of them worth further exploration in order to better treat the relativistic effects under different circumstances.

\section{Conclusion}
\label{section:conclusion}
In this paper, we have extensively tested the performance of some mainstream ECPs under the FermiNet for the first two row elements. Based on the results, we found that the qualities and features of the ECPs have been correctly maintained under the FermiNet. Two more recently built ECP tables, namely ccECP and eCEPP, prevail in the contest with errors on average below chemical accuracy. And ccECP performs slightly better on the spectral precision and covers more elements, while eCEPP is more systematically built from both shape and energy consistency, and better treats the core polarisation effects. Besides, even though the all-electron results are still very accurate for the light elements, the errors start to overwhelm due to a lack of relativistic treatment as we get to the second row. We also observe that the all-electron calculations of some heavier elements are likely prone to the numerical instabilities. Finally, based on further discussions, we have generated few future directions for developing and improving the FermiNet.

\begin{acknowledgments}
The work of Mengsa Wang and Han Wang is supported by the National Key R\&D Program of China (Grant No.~2022YFA1004300) and the National Natural Science Foundation of China (Grant No.~12122103). 
Y. Zhou's work is supported by National Key R {\&} D Program of China under grants 2019YFA0709600 and 2019YFA0709601 and National Natural Science Foundation of China under grant 12004047.
\end{acknowledgments}

\appendix
\renewcommand\thesubsection{\Alph{subsection}}
\renewcommand\theequation{\thesubsection.\arabic{equation}}
\section*{Appendix}
\counterwithin*{equation}{subsection}

\subsection{Hyperparameters}
\label{app:hyper}
The hyperparameters used in this work are given in Table \ref{table:11}. The learning rate is gradually decaying with the training step $t$:
\begin{equation}
    \text{learning rate} =\frac{lr_{0}}{1+\frac{t}{t_{0}}},
\end{equation}
where $lr_{0}$ is the initial learning rate and $t_{0}$ is also known as the hyperparameter $lr.delay$, which controls the scale of rate decay. Large learning rate may induces instability in the optimization process, causing the hyperparameters to fluctuate rather than smoothly converging to the optimal solution during training. We observed that for Sulfur and Chlorine, as well as their cations and anions, using the default learning rate causes numerical instabilities in the training processes, so we chose to reduce the learning rate by tuning down the hyperparameter $lr.delay$.

\begin{table}[htbp]
    \centering
    \caption{\small The hyperparameter values used for the calculations in this paper.}
    %\footnotesize
	
	~\\
	
    \small
    \begin{tabular}{ c c c }
		\hline
		\hline
		&Parameter&Value\\
		\hline
		\multicolumn{1}{c}{\multirow{2}*{Pretrain}}&basis set&cc-pVDZ\\
		~&iterations&10000\\
		\hline
		\multicolumn{1}{c}{\multirow{4}*{Training}}&iterations&100000\\
		~&initial learning rate&0.05\\
		~&learning rate delay&10000\\
		~&optimizer&KFAC\\
		~&local energy clipping&5.0\\
		\hline
		Inference&iterations&10000\\
		\hline
		\multicolumn{1}{c}{\multirow{4}*{MCMC}}&batch size&4096\\
		~&mcmc steps&10\\
		~&init width&1.0\\
		~&move width&0.02\\
		\hline
		\multicolumn{1}{c}{\multirow{5}*{Networks}}&hidden units&(256,32)\\
		~&layer&4\\
		~&determinants&16\\
		~&precision&double\\
		~&envelope&isotropic\\
            ~&full\_det&true \\
		~&ecp quadrature& 12 points icosahedron\\
		\hline
		\hline
    \end{tabular}

    \label{table:11}
\end{table}

\subsection{Efficiency}	
\label{app:eff}
The ECP scheme inevitably introduces additional costs. Here, we assess them in detail. The expression for calculating the local energy with the effective core potential is as follows:
\begin{equation}
    E_{\text{L}}(\textbf{r})=\psi^{-1}(\textbf{r})\hat{H}\psi(\textbf{r})+E_{\text{nl}}(\textbf{r}).	
\end{equation}
We employed a 12-point icosahedral quadrature method to approximate the energy of the nonlocal term, as done in Refs.~\cite{li2022fermionic,qmc}. The specific form of the nonlocal energy component $E_{\text{nl}}(\textbf{r})$ is given by:
\begin{equation}
    \begin{split}
		E_{\text{nl}}(\textbf{r})&=\sum_{i}E_{\text{nl},i}(\textbf{r}_{i})=\sum_{i}\sum_{l}V_{l}(r_{i})\sum_{m=-l}^{l}Y_{lm}(\Omega_{i})\\&\int Y_{lm}^{*}(\Omega_{i^{'}})\frac{\psi(\textbf{r}_{1},\textbf{r}_{2},...,\textbf{r}_{i-1},\textbf{r}_{i}^{'},...,\textbf{r}_{N_{v}})}{\psi(\textbf{r}_{1},\textbf{r}_{2},...,\textbf{r}_{i-1},\textbf{r}_{i},...,\textbf{r}_{N_{v}})}d\Omega_{i^{'}}.
    \end{split}	
\end{equation}

\begin{figure}[htp]
    \centering
    \includegraphics[scale=0.12]{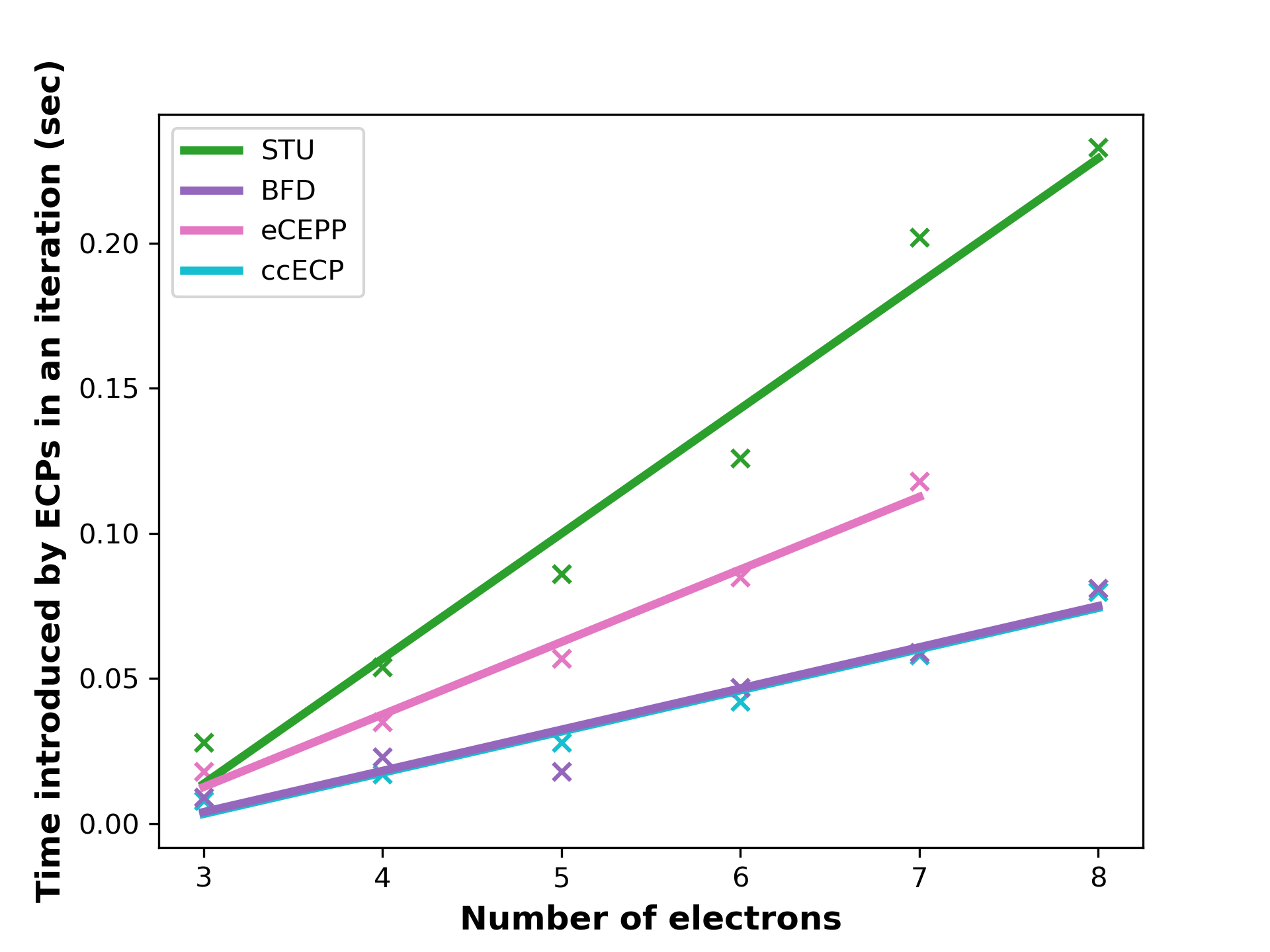}
    \caption{\small Extra runtime introduced by different ECPs for elements from B to Ne. }
    \label{picture:4}
    \end{figure}

The number of electrons and the averaged runtime of a single training iteration for the all-electron and ECPs calculations on 8 Nvidia A100 GPUs is shown in Table~\ref{table:10}. We studied the asymptotic behavior of the extra time with respect to the number of electrons in the single-atom calculations from B to Ne, as shown in Fig.~\ref{picture:4}. We found that the extra runtime increases linearly with the number of electrons for fixed $l_{\text{max}}$, suggesting that the time of computing single numerical integration is a fixed constant, which we denote as $C_{\text{int}}$. The value of $l_{\text{max}}$ determines the number of integrals to be calculated for one electron. The effect of the number of Gaussian functions in calculations can be neglected. Therefore, we give a quantitative estimate of the additional computation brought by the ECP scheme, which is $\mathcal{O}(N_{v}(l_{\text{max}}+1)C_{\text{int}})$.

\begin{figure}[htp]
    \centering
    \includegraphics[scale=0.088]{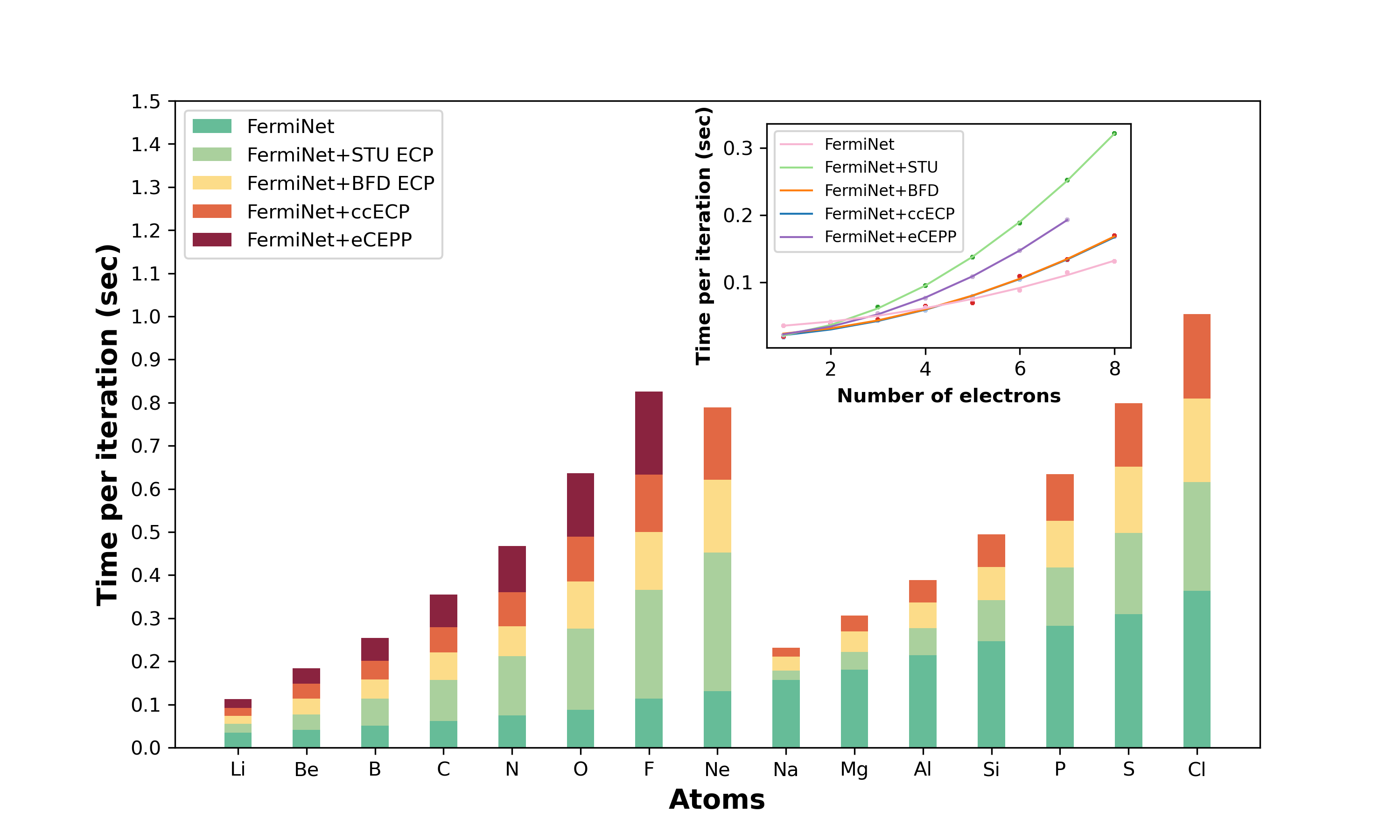}
    \caption{\small Runtime of a single training iteration for FermiNet with different ECPs and AE for elements from Li to Cl.}
    \label{picture:5}
\end{figure}

We also investigated the asymptotics  behavior of FermiNet with ECPs. Using quadratic, cubic and quartic polynomials, respectively, to fit the curves for the computational complexity of the first row of elements, we found that the quadratic polynomials are able to fit the curves well in Fig.~\ref{picture:5}. This suggests that the complexity of FermiNet with ECPs scale as $\mathcal{O}(N_{v}^{2})$, where the computation of the single-electron stream becomes dominant. And the computational gain starts from the second row of elements in the ECP scheme. 
	
\subsection{Envelope}
\label{app:env}
To ensure that the wavefunction converges to zero at infinity, FermiNet adopts different envelope functions to shape the function $\phi_{i}^{k\alpha}(\textbf{r}_{j}^{\alpha};\{\textbf{r}_{/j}^{\alpha}\};\{\textbf{r}^{\overline{\alpha}}\})$:
\begin{equation}
    \begin{split}
    &\phi_{i}^{k\alpha}(\textbf{r}_{j}^{\alpha};\{\textbf{r}_{/j}^{\alpha}\};\{\textbf{r}^{\overline{\alpha}}\})\\&=(\textbf{w}_{i}^{k\alpha}\cdot \textbf{h}_{j}^{L\alpha}+g_{i}^{k\alpha})\sum_{m}\pi_{im}^{k\alpha}\exp\left(-\left|\Sigma_{im}^{k\alpha}(\textbf{r}_{j}^{\alpha}-\textbf{R}_{m})\right|\right).
    \end{split}
\end{equation}
When using the parameters $\Sigma_{im}^{k\alpha}\in \mathbb{R}^{3\times 3}$, we refer to it as the full envelope. In order to simplify the network and reduce the number of parameters, FermiNet\cite{spencer2020better} replaces $\Sigma_{im}^{k\alpha}\in \mathbb{R}^{3\times 3}$ with $\sigma_{im}^{k\alpha}\in \mathbb{R}$. The latter can be understood as a diagonal matrix with identical diagonal elements, ensuring that the wavefunction is isotropically converged in space, commonly known as the isotropic envelope.
	
\begin{figure}[htp]
    \centering
    \includegraphics[scale=0.5]{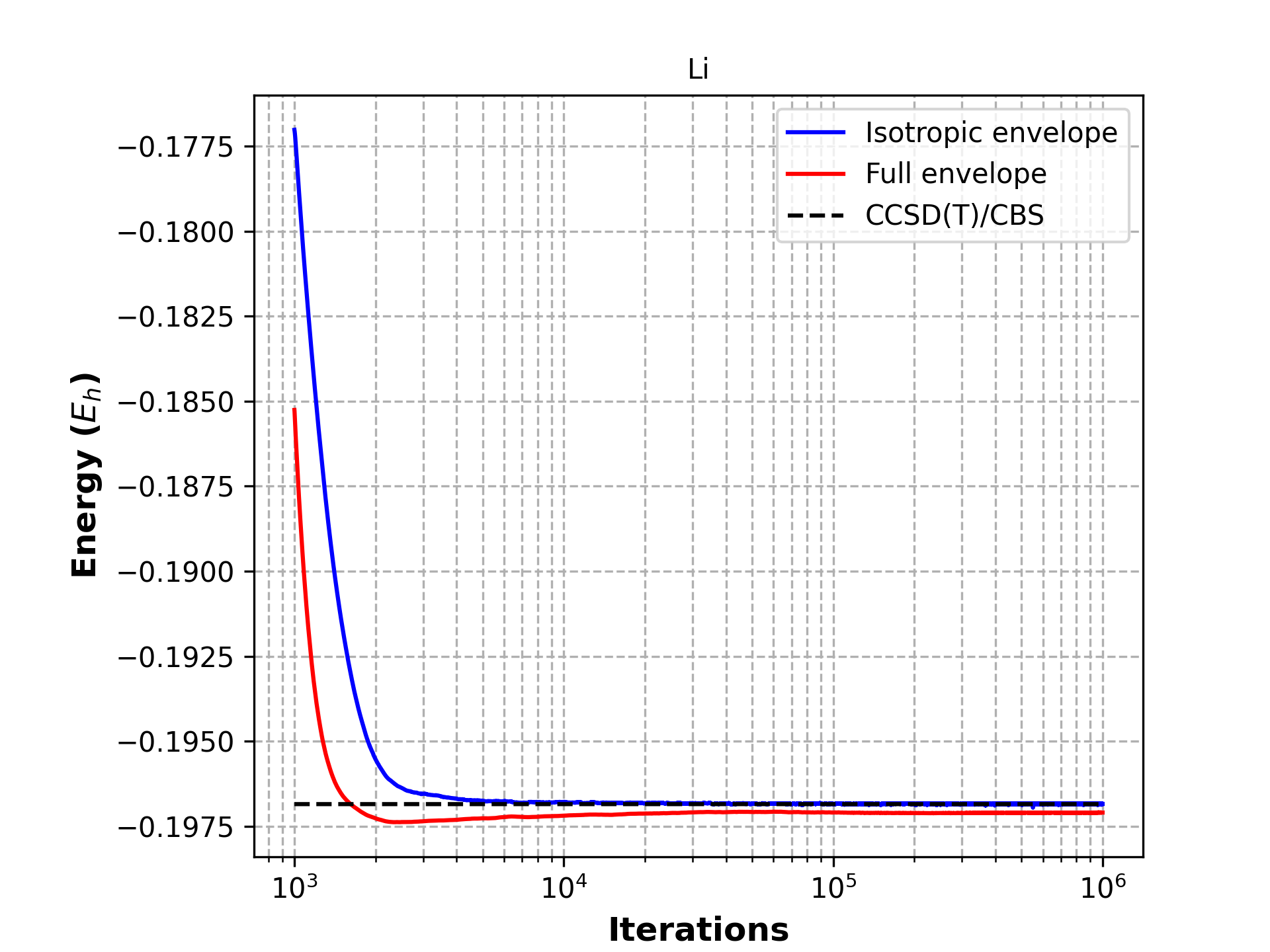}
    \caption{\small Learning curves of the ground state energy of Li using full and isotropic envelopes, respectively. Energies are smoothed over 1000 iterations.}
    \label{picture:6}
\end{figure}

For Li, we calculated the ground state energy of FermiNet with ccECP using the full envelope and isotropic envelope, respectively. In these tests, we pretrain the wavefunction by $1,000$ iterations and train the wavefunction with $1,000,000$ iterations and single-precision. Here, we utilized the code from Ref.~\cite{li2022fermionic} to calculate the case with the full envelope. In Table \ref{table:6} and Fig.~\ref{picture:6}, the results showed that the ground state energy using the full envelope is lower than the result of the isotropic envelope and even lower than the result obtained by CCSD(T)/CBS+ccECP\cite{annaberdiyev2020accurate}. Yet considering the symmetry of the atomic systems, we believe that it is more reasonable to use the isotropic envelope. Therefore, the isotropic envelope is used in this work.
	
\begin{table}[htpb]
    \centering
    \caption{\small Ground state energy of Li calculated by FermiNet with ccECP using the full envelope and the isotropic envelope, respectively.}
    %\scriptsize
    \small
    ~\\
    \begin{tabular}{ c@{\hspace{0.3cm}}  c@{\hspace{0.3cm}}  c @{\hspace{0.3cm}} c@{\hspace{0.3cm}}  c}
	\hline
		\hline
		&\multicolumn{2}{c}{CCSD(T)+ccECP\cite{annaberdiyev2020accurate}}&\multicolumn{2}{c}{FermiNet+ccECP}\\
		\cmidrule(lr){2-3}\cmidrule(lr){4-5}&5Z&CBS&Isotropic&Full\\
		\cmidrule(lr){2-3}\cmidrule(lr){4-5}
		Li&-0.19685279&-0.19685279(1)&-0.196844(2)&-0.1970996(1)\\
		\hline
		\hline
    \end{tabular}
    \label{table:6}
\end{table}
	
\subsection{Precision}
\label{app:prec}	
In our numerical experiments, we found the ground state energy obtained from all-electron calculation using single-precision is even lower than the exact energy for Nitrogen \cite{chakravorty1993ground}. We speculated that this weird phenomenon is caused by numerical errors in the calculations. For the ground state energy of N, the effective precision in single-precision is approximately 5 to 6 decimal places. As the training progresses, the accumulated numerical errors might go beyond this threshold. However, in some situations, such as the investigation of the fine structure of atoms and molecules, as well as the high-precision quantum chemical calculations, chemical accuracy may prove inadequate, prompting the necessity for a higher level of precision. Consequently, double-precision is used to guarantee the accuracy.
	
\begin{figure}[htp]
    \centering
    \includegraphics[scale=0.21]{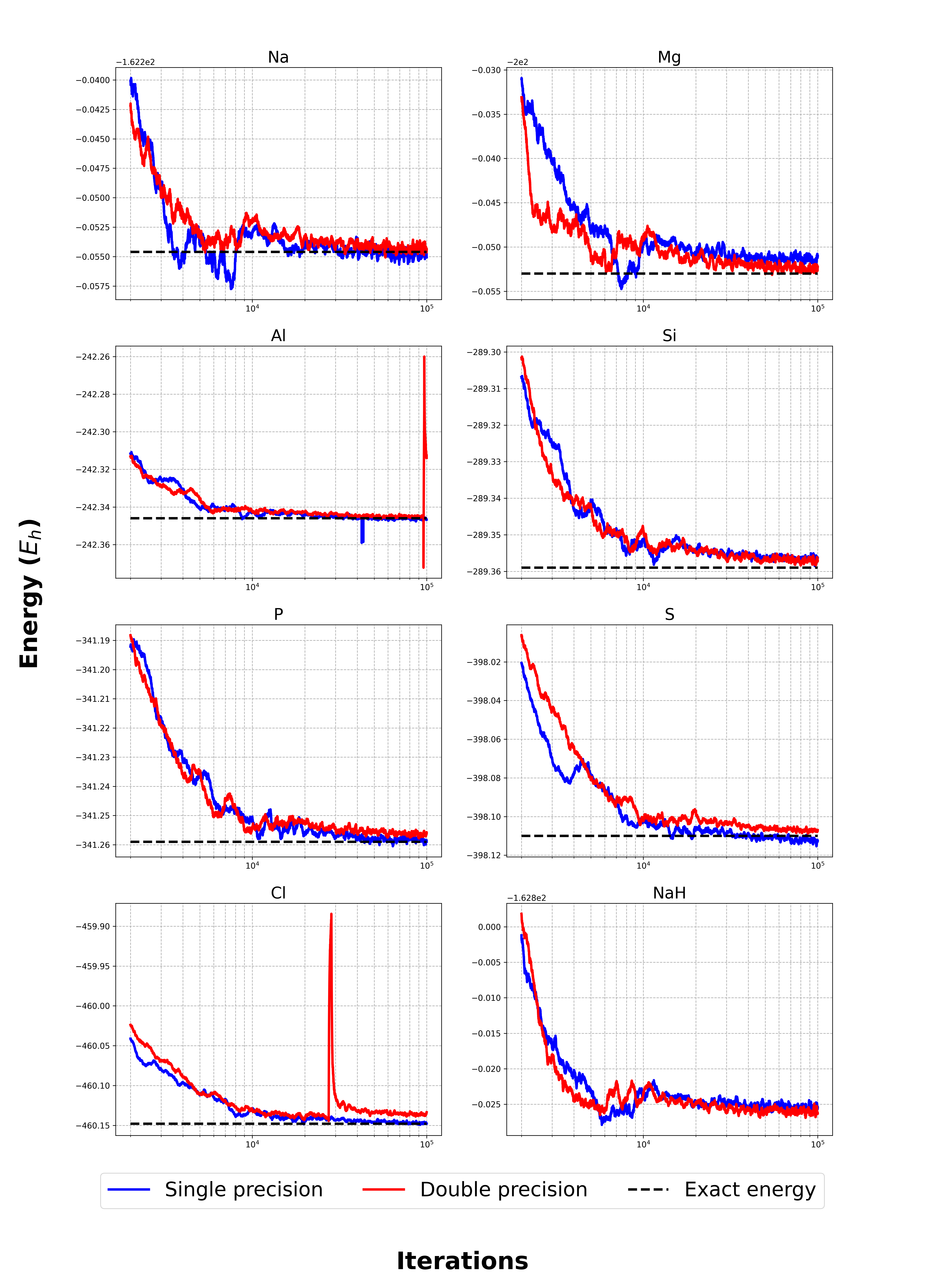}
    \caption{\small Learning curves of the all-electron ground state energies for elements from Na to Cl and NaH with single-precision and double-precision, respectively. In this figure, all parameters are default values. That is, the parameter $lr.delay$ is set to $10,000$. Energies are smoothed over 1000 iterations.}
    \label{picture:7}
\end{figure}
	
Although double-precision can guarantee the accuracy, it is observed that the training process using double-precision is not as stable as single-precision. In Figure \ref{picture:7}, When employing double-precision and setting $lr.delay$ to $10,000$ for calculating the all-electron ground state energy of Aluminum, an instability is observed, leading to an explosion in the calculation around training steps of $100,000$, and the same issue occurs during the training process for Sulphur and Chlorine and the corresponding cations and anions. We solved this issue by reducing $lr.delay$ to $5,000$ or $3,000$.	

\begin{figure}[htp]
    \centering
    \includegraphics[scale=0.48]{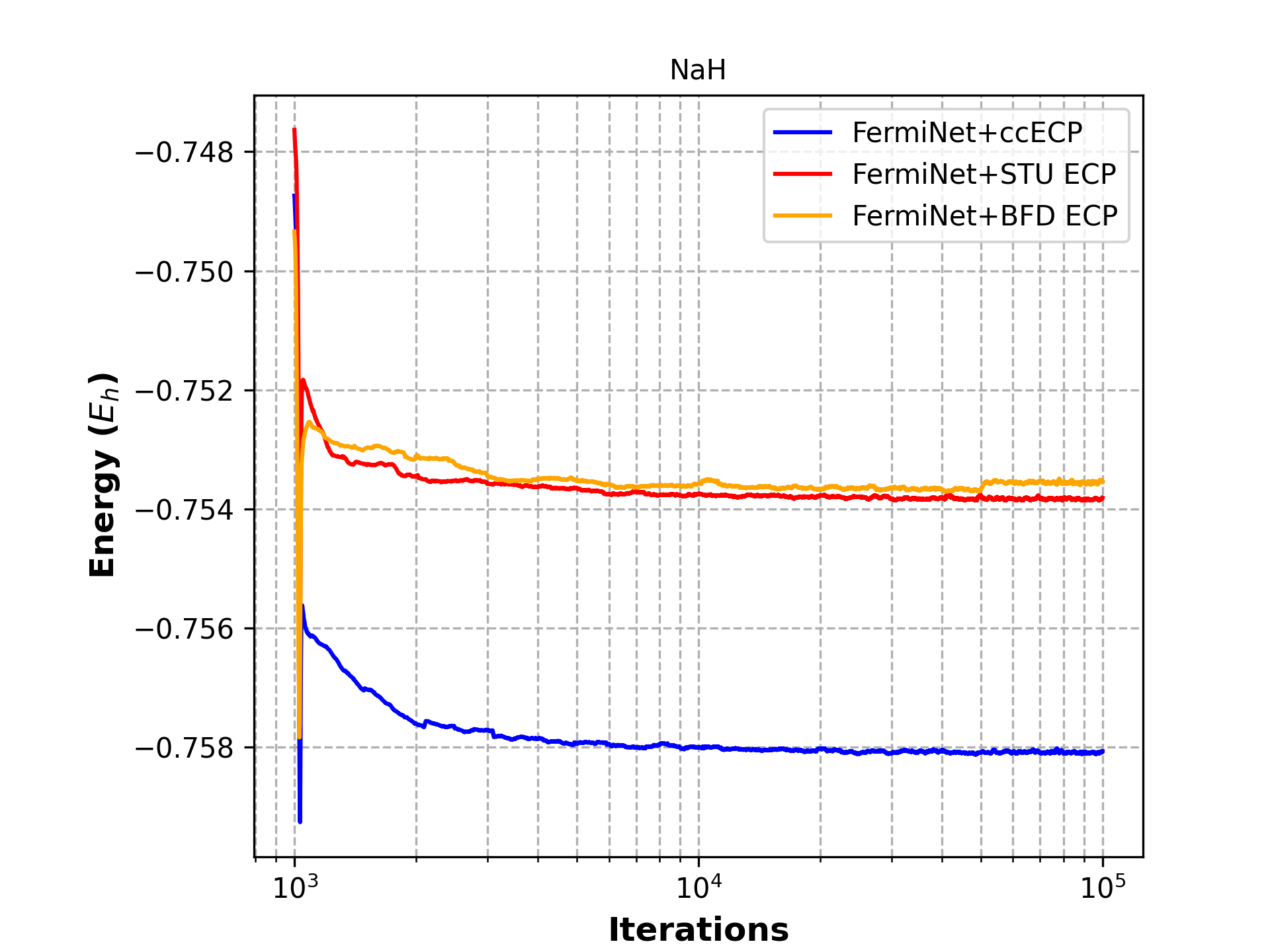}
    \caption{Learning curves of NaH  with the STU, BFD and ccECP, respectively.}
    \label{picture:8}
\end{figure}

When computing the ground state energy of NaH with various ECPs, we observed a sharp change at the beginning of training in Fig.~\ref{picture:8}. It might be attributed to the substantial difference of wavefunction between HF and FermiNet, leading to a significant deviation in sampling during the early stage of training. As the training proceeds, the ground state energy gradually stabilizes and eventually converges to a reasonably value.

\subsection{Estimation of relativistic effects}
\label{appendix1}
In order to better compare between the all-electron results and the experimental data, we have incorporated the approximated relativistic effects into the all-electron results. These effects are obtained from Ref.\cite{koga1997atomic} and Ref.\cite{klopper2010sub}, which are combinations of literature results, analytical and perturbative calculations at CCSD(T) and HF levels. The corrected all-electrons IP and EA are shown in Table \ref{table:7}. Specifically, for the first row of elements, we obtained the relativistic all-electron IP and EA  by adding the sum of scalar relativistic contributions, the spin-orbit correction and diagonal Born-Oppenheimer correction, consistent with the original FermiNet paper \cite{ferminet}. For the second row of elements, we obtained the relativistic IP and EA by adding the sum of mass, LS-non-splitting and fine-structure corrections. 

In Table \ref{table:x2c}, we provide the estimation of relativistic correlation calculated from X2C\cite{X2C} at the HF level using PySCF\cite{sun2018pyscf}, and compare with existing relativistic corrections. We also calculate relativistic corrections on DE for some selected hydrides in the same manner, as shown in Table~\ref{table:x2c_DE}.
Specifically, we use ccpv5z-dk basis set for the X2C level calculations, and ccpv5z basis set for the non-relativistic calculations.

Finally, we note that both schemes have significant drawbacks. The existing one, though might be accurate for those from analytical and perturbative CCSD(T) calculations, is apparently unsystematic. While X2C corrections at HF level, though being more systematic, suffer from the inaccuracies from HF calculations. For instance, the HF predicted EA of many elements differ in sign with those from FermiNet or CCSD(T). As a result, one may also question the accuracy of X2C corrections. Yet these numbers should be correct on magnitude and sufficient for the purpose of giving a feeling on the scale of relativistic effects. 

%The above two estimates of relativistic effects are both rough, with relatively large deviations from experimental values in the calculations of electron affinity for certain elements, sometimes even differing in sign. Therefore, systematically and accurately calculating relativistic effects remains an important and challenging issue that must be tackled

\begin{table}[htpb]
    \centering
    \small
    \caption{\small{IP and EA from the all-electron calculations with the relativistic effects for atoms from Li to Cl. }}
~\\
    \begin{threeparttable}
    \begin{tabular}{ c@{\hspace{0.32cm}}  c@{\hspace{0.32cm}}  c@{\hspace{0.32cm}}  c@{\hspace{0.32cm}}  c@{\hspace{0.32cm}}  c @{\hspace{0.32cm}} c }
	\hline
		\hline
		&\multicolumn{3}{c}{Ionization Potential($mE_{h}$)}&\multicolumn{3}{c}{Electron affinity($mE_{h}$)}\\ 
		%\cline{2-9}
		\cmidrule(lr){2-4}\cmidrule(lr){5-7}
		Atom&FermiNet$^\dagger$\tnote{a,b}&Expt.&$\Delta E$&FermiNet$^\dagger$\tnote{c,d}&Expt.&$\Delta E$\\ 
		\hline
        Li&197.91\tnote{a}&198.14&-0.23&21.48\tnote{c}&22.71&-1.23\\
	Be&342.57\tnote{a}&342.60&-0.03 &-&-&-\\  
	B&304.88\tnote{a}&304.95& -0.07&9.53\tnote{c}&10.18&-0.65\\
	C&413.67\tnote{a}&413.81& -0.14&46.34\tnote{c}&46.41&-0.07\\
	N&534.01\tnote{a}&534.12&-0.11&-&-&-\\
        O&500.25\tnote{a}&500.45&-0.2&53.10\tnote{c}&53.69&-0.59\\
	F&640.06\tnote{a}&640.28&-0.22 &125.25\tnote{c}&124.99&0.26\\
	Ne&792.52\tnote{a}&792.48&0.04&-&-&\\
    Na&188.51\tnote{b}&188.86&-0.35&20.88\tnote{d}&20.14&0.74\\
	Mg&280.66\tnote{b}&280.99&-0.33&-&-&-\\
	Al&219.01\tnote{b}&219.97 &-0.96&14.57\tnote{d}&16.21 &-1.64\\
		Si&298.74\tnote{b}&299.57&-0.83&52.02\tnote{d}&50.90&1.12\\
	P&385.23\tnote{b}&385.38&-0.15 &24.08\tnote{d}&27.43&-3.35\\
		S&380.31\tnote{b}&380.72&-0.41&73.66\tnote{d}&76.33&-2.67\\
		Cl&474.08\tnote{b}&476.55&-2.47&132.20\tnote{d}&132.76&-0.56\\
		\hline
		MAD$_{1}$&&&0.13&&&0.56\\
        MAD$_{2}$&&&0.79&&&1.68\\
        MAD&&&0.44&&&1.17\\
		\hline
		\hline
    \end{tabular}
    \begin{tablenotes} 
        \footnotesize
        \item[a] Relativistic effects in the IPs from Li to Ne are from Ref.\cite{klopper2010sub}.
        $\text{FermiNet}^\dagger\text{.IP}=\text{FermiNet.IP}+\text{E}_{\text{Rel.}}+\text{E}_{\text{SO}}+\text{E}_{\text{DBOC}}.$
        \item[b] Relativistic effects in the IPs from from Na to Cl are from Ref.\cite{koga1997atomic}.
        $\text{FermiNet}^\dagger\text{.IP}=\text{FermiNet.IP}+\text{E}_{\text{LS-non-split}}+\text{E}_{\text{fine-struc}}+\text{E}_{\text{mass correc}}.$
        \item[c] Relativistic effects in the EAs from Li to Ne are from Ref.\cite{klopper2010sub}.
        $\text{FermiNet}^\dagger\text{.EA}=\text{FermiNet.EA}+\text{E}_{\text{Rel.}}+\text{E}_{\text{SO}}+\text{E}_{\text{DBOC}}.$
        \item[d] Relativistic effects in the EAs from from Na to Cl are from Ref.\cite{koga1997atomic}.
        $\text{FermiNet}^\dagger\text{.EA}=\text{FermiNet.EA}+\text{E}_{\text{LS-non-split}}+\text{E}_{\text{fine-struc}}+\text{E}_{\text{mass correc}}.$
    \end{tablenotes}
    \end{threeparttable}
    \label{table:7}
\end{table}

\begin{table}[htpb]
    \centering
    \small
    %\footnotesize
    \caption{\small{Estimations of the relativistic effects of ionization potential and electron affinity from Li to Cl used by this work (obtained from Refs.~\cite{klopper2010sub,koga1997atomic}) and those by X2C at the HF level. }}
~\\
    \begin{threeparttable}
    \begin{tabular}{ c@{\hspace{0.3cm}}  c@{\hspace{0.3cm}}  c@{\hspace{0.3cm}}  c@{\hspace{0.3cm}} c}
	\hline
		\hline
		&\multicolumn{2}{c}{Ionization Potential($mE_{h}$)}&\multicolumn{2}{c}{Electron affinity($mE_{h}$)}\\ 
		%\cline{2-9}
		\cmidrule(lr){2-3}\cmidrule(lr){4-5}
		Atom&X2C\tnote{a}&this work\tnote{b,c}&X2C\tnote{a}&this work\tnote{b,c}\\ 
		\hline
        Li&0.014\tnote{a}&-0.005\tnote{b}&-0.00002\tnote{a}&-0.003\tnote{b}\\
	Be&0.042\tnote{a}&0.010\tnote{b}&-&-\\  
	B&-0.088\tnote{a}&-0.032\tnote{b}&-0.058\tnote{a}&-0.057\tnote{b}\\	 
        C&-0.156\tnote{a}&-0.206\tnote{b}&-0.123\tnote{a}&-0.228\tnote{b}\\
	N&-0.255\tnote{a}&-0.659\tnote{b}&-&-\\
        O&-0.372\tnote{a}&0.0004\tnote{b}&-0.305\tnote{a}&-0.298\tnote{b}\\
	F&-0.535\tnote{a}&-0.673\tnote{b}&-0.467\tnote{a}&-0.968\tnote{b}\\
	Ne&-0.745\tnote{a}&-1.926\tnote{b}&-&-\\
        Na&0.224\tnote{a}&0.218\tnote{c}&0.005\tnote{a}&-0.004\tnote{c}\\
	Mg&0.355\tnote{a}&0.343\tnote{c}&-&-\\
	Al&-0.33\tnote{a}&-0.054\tnote{c} &-0.196\tnote{a}&-0.317\tnote{c}\\
        Si&-0.413\tnote{a}&-0.637\tnote{c}&-0.348\tnote{a}&-0.955\tnote{c}\\
	P&-0.503\tnote{a}&-1.883\tnote{c}&-0.421\tnote{a}&0.035\tnote{c}\\
		S&-0.58\tnote{a}&0.238\tnote{gc}&-0.546\tnote{a}&-0.697\tnote{c}\\
		Cl&-0.719\tnote{a}&-0.969\tnote{c}&-0.689\tnote{a}&-1.995\tnote{c}\\
		\hline
		\hline
    \end{tabular}
    \begin{tablenotes} 
        \footnotesize
        \item[a] Estimation of relativistic effects on ionization potential and electron affinity, derived from X2C at the HF level using ccpv5z-dk basis sets, for elements ranging from Li to Cl.
        \item[b] Estimation of relativistic effects on ionization potential and electron affinity for elements from Li to Ne from Ref.\cite{klopper2010sub}.
        $\text{Relativistic Correction}=\text{E}_{\text{Rel.}}+\text{E}_{\text{SO}}+\text{E}_{\text{DBOC}}.$
        
        \item[c]Estimation of relativistic effects on ionization potential and electron affinity for elements from Na to Cl from Ref.\cite{koga1997atomic}.
        $\text{Relativistic Correction}=\text{E}_{\text{LS-non-split}}+\text{E}_{\text{fine-struc}}+\text{E}_{\text{mass correc}}.$
    \end{tablenotes}
    \end{threeparttable}
    \label{table:x2c}
\end{table}

\begin{table}[htpb]
    \centering
    \small
    %\footnotesize
    \caption{\small{Estimations of relativistic effects  on DE used by this work (obtained from ref.\cite{visscher1996relativistic,peebles2002high}) and those calculated by X2C at the HF level using ccpv5z-dk basis sets for HF, H$_2$S and HCl.}}
~\\
    \begin{tabular}{ c@{\hspace{1.cm}}  c@{\hspace{1.0cm}}  c@{\hspace{1.0cm}}  c}
	\hline
		\hline
        &\multicolumn{3}{c}{Relativistic effect($mE_{h}$)}\\
        \cmidrule(lr){2-4}
		&HF\tnote{e}&H$_2$S&HCl\\ 
		\hline
         this work~\cite{visscher1996relativistic,peebles2002high}&-0.96\cite{visscher1996relativistic}&-1.71\cite{peebles2002high}&-1.91\cite{visscher1996relativistic}\\
        X2C&-0.40&-0.83&-0.52\\
		\hline
		\hline
    \end{tabular}
    \label{table:x2c_DE}
\end{table}

\subsection{Other supporting Data}
\label{appendix2}

\begin{figure}[htp]
    \centering
    \includegraphics[scale=0.34]{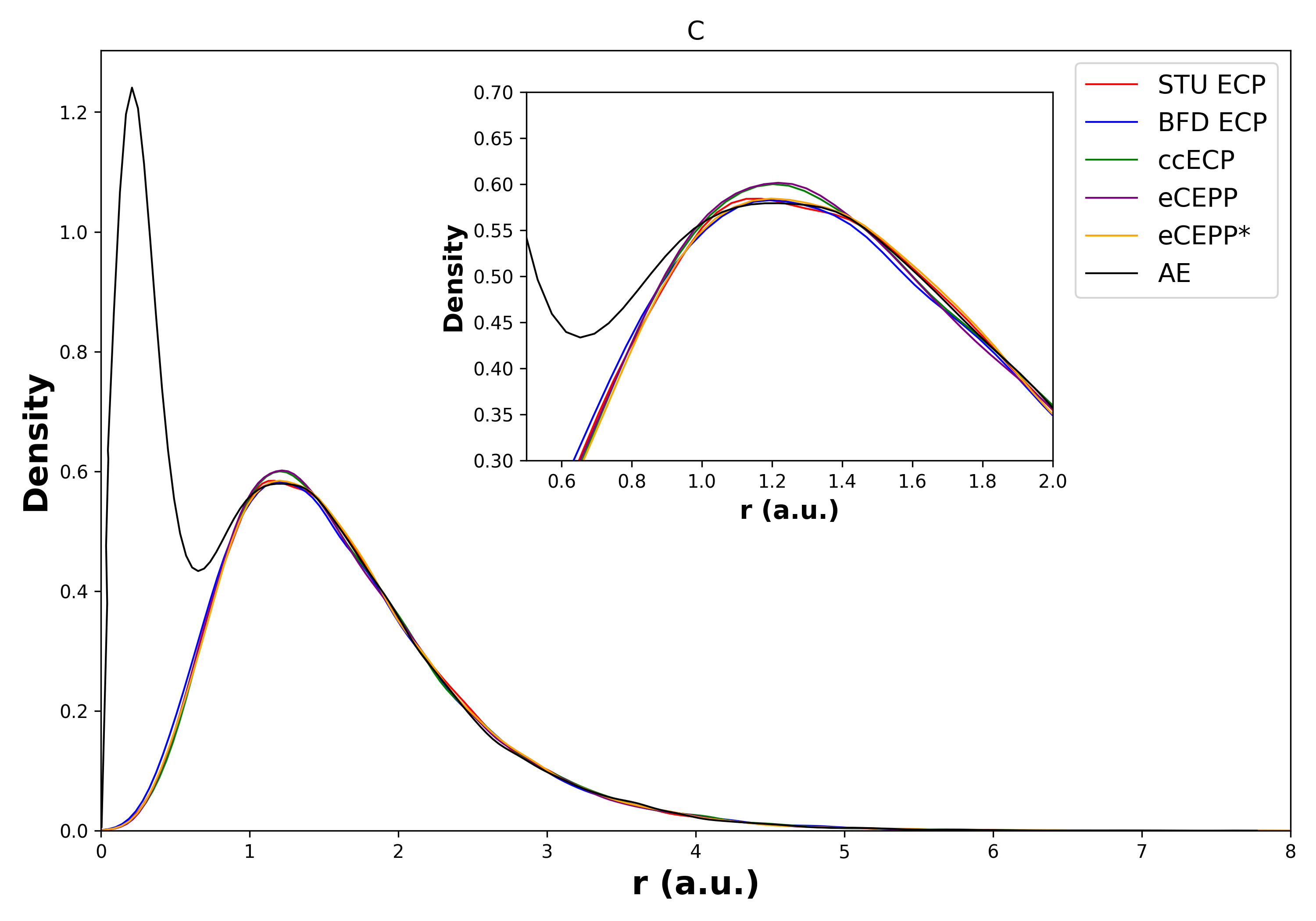}
    \caption{The radial density distribution from the ground state Carbon atom.}
    \label{picture:density_C}
\end{figure}

\begin{figure}[htp]
    \centering
    \includegraphics[scale=0.34]{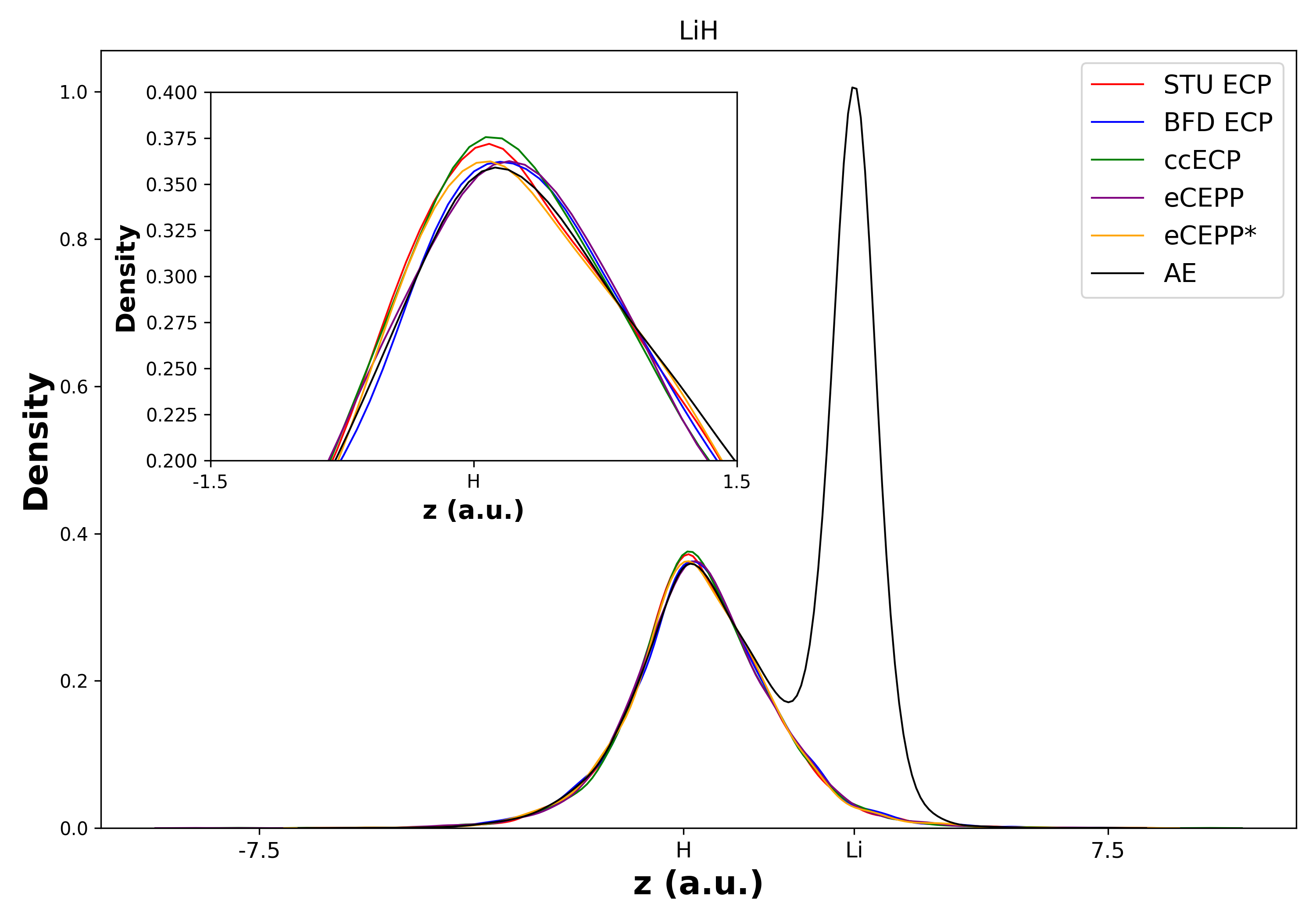}
    \caption{The axial density distributions along the z-axis of the ground state wavefunctions of LiH.}
    \label{picture:density_LiH}
\end{figure}
More detail results have been placed here to better support the arguments in the main text. In Fig.~\ref{picture:density_C} and Fig.~\ref{picture:density_LiH}, we illustrate the radial density distributions of carbon and the axial density distributions of LiH. In Table \ref{table:8}, we provide the ground state energies of atoms from the ECP calculations. In Table \ref{table:9}, we provide the ground state energies of the hydrides.

In Table \ref{table:valcorr}, we report the valence correlation energies from different ECPs for ground state energy, IP and EA. These values are obtained by subtracting the HF energy from FermiNet's energy. The all-electron valence correlation energies are from CCSD(T) calculations and serve as reference \cite{ranasinghe2015core}. We calculated the HF energy via PySCF package\cite{sun2018pyscf}. For STU ECP, we used the Stuttgart basis set. For the other three ECPs, we used the corresponding valence basis sets and performed calculations with $n$ = 5. And to ensure fairness, we omitted the CPP term from the eCEPP calculations. 
%and extrapolated to the CBS limit with the following formula,
%\begin{equation}    \text{E}_{n}^{\text{HF}}=\text{E}_{\text{CBS}}^{\text{HF}}+a\text{exp}^{-bn},\end{equation}
%where $n$ labels the basis set size, $a$ and $b$ are both the fitting parameters. If the CBS limit is not possible, we used the results from the $n$ = 5 calculations. And to ensure fairness, we omitted the CPP term from the eCEPP calculations.}

In Table \ref{table:10}, we summarize the number of electrons and the single iteration runtime for the all-electron and various ECPs calculations on 8 Nvidia A100 GPUs. They have been averaged over 1000 training iterations. In Table~\ref{table:refFerminet}, we give the results of ground state energies on the atoms and CH$_4$ from the reference Ferminet.

\begin{table*}
    \centering
    %\small
    %\scriptsize
    \footnotesize 
    \caption{\small{Ground state energies from Li to Cl with different ECPs.}}	
		~\\	
    \begin{tabular}{ c@{\hspace{0.4cm}} c@{\hspace{0.4cm}} c@{\hspace{0.4cm}} c@{\hspace{0.4cm}} c@{\hspace{0.4cm}} c}
		\hline
		\hline
		&\multicolumn{5}{c}{Ground state energy$(E_{h})$}\\ 
		%\cline{2-9}
		\cmidrule(lr){2-6}
		Element&STU&BFD&eCEPP&eCEPP*&ccECP\\ 
		\hline
		Li&-0.19602&-0.19631&-0.19783 &-0.19585&-0.19683\\
		Be&-1.01095&-1.00969&-1.01315&-1.00770&-1.01011\\  
		B&-2.61904&-2.62016&-2.62539&-2.61552&-2.61519\\
		C&-5.43345&-5.43423&-5.43435&-5.41993&-5.41737\\
		N&-9.80063&-9.80235&-9.78748&-9.76716&-9.76586\\
		O&-15.90626&-15.90669&-15.86207&-15.84674&-15.88384\\
		F &-24.19913&-24.20082&-24.16549&-24.14957&-24.19683\\
		Ne&-35.03760&-35.03840&-&-&-35.03438\\
		Na&-0.18210&-0.18210&-&-&-0.18618\\
		Mg&-0.82068&-0.81974&-&-&-0.82346\\
	Al&-1.93931&-1.93913&-&-&-1.93727\\
		Si&-3.76770&-3.76843&-&-&-3.76180\\
		P&-6.47667&-6.47882&-&-&-6.45953\\
		S&-10.13308&-10.13618&-&-&-10.09687\\
		Cl&-14.97826&-14.97868&-&-&-14.92612\\
		\hline
		\hline
    \end{tabular}
    \label{table:8}
\end{table*}

\begin{table*}
    \centering
    %\scriptsize
    %\small
    \footnotesize 
    \caption{\small{Ground state energies of the hydrides from the all-electron and ECPs calculations. }}
    ~\\
    \begin{tabular}{ c@{\hspace{0.4cm}} c @{\hspace{0.4cm}}c@{\hspace{0.4cm}} c@{\hspace{0.4cm}} c@{\hspace{0.4cm}} c@{\hspace{0.4cm}} c }
		\hline
		\hline
		&\multicolumn{6}{c}{Ground state energy ($E_{h}$)} \\ 
		%\cline{2-9}
		\cmidrule(lr){2-7}
		Element&AE&STU&BFD&eCEPP&eCEPP*&ccECP\\ 
		\hline
		LiH &-8.07051&-0.78756 &-0.78819&-0.79120&-0.78699&-0.78819\\
		BeH$_2$&-15.90333&-2.24743&-2.24542&-2.25058&-2.24151&-2.24631\\  
		BH$_3$&-26.60170&-4.56505&-4.56874&-4.57546&-4.56158&-4.56237\\
		CH$_4$&-40.51410 &-8.09934&-8.10065&-8.10557&8.08707&-8.08728\\
		NH$_3$ &-56.56337&-11.77127 &-11.77366&-11.76236&-11.73973&-11.74110\\
		H$_2$O &-76.43786&-17.27501&-17.27592&-17.23355&-17.21697&-17.25388\\
		HF&-100.45902&-24.92350&-24.92554&-24.89074&-24.87425&-24.92198 \\
		NaH&-162.8261&-0.75388&-0.75352&-&-&-0.75855\\
		MgH&-200.60296&-1.37381&-1.37369 &-&-&-1.37559\\
		AlH&-242.96184&-2.55740&-2.55686&-&-&-2.55455\\
		SiH$_4$&-291.87121&-6.28568&-6.28620&-&-&-6.27826\\
		PH$_3$&-343.138&-8.35769&-8.36300&-&-&-8.34428\\
		H$_2$S&-399.39499&-11.42607&-11.42762&-&-&-11.38902\\
		HCl&-460.81123&-15.64912&-15.65035&-&-&-15.59698\\
		\hline
		\hline
    \end{tabular}
    \label{table:9}
\end{table*}

%\subsection{Efficiency}
%\label{appendix3}

\begin{table*}[htpb]
    \centering
    %\scriptsize
    %\small
    \footnotesize 
    \caption{\small Valence correlation energy from the all-electron and ECPs calculations.}
    ~\\
    \begin{tabular}{ c c c c c c c c c c c c c c c c }
		\hline
		\hline
		&\multicolumn{5}{c}{Ground state ($mE_{h}$)}&\multicolumn{5}{c}{Ionization Potential ($mE_{h}$)}&\multicolumn{5}{c}{Electron affinity ($mE_{h}$)} \\ 
		%\cline{2-9}
		\cmidrule(lr){2-6}\cmidrule(lr){7-11}\cmidrule(lr){12-16}
		Element&AE\cite{ranasinghe2015core}&STU&BFD&ccECP&eCEPP*&AE\cite{ranasinghe2015core}&STU&BFD&ccECP&eCEPP*&AE\cite{ranasinghe2015core}&STU&BFD&ccECP&eCEPP*\\ 
		\hline
Be&-46.12&-48.98&-47.75&-48.22&-48.02&-&-&-&-&-&-&-&-&-&-\\										
B&-72.94&-78.33&-76.17&-75.63&-75.53&11.61&5.04&9.99&10.20&9.75&19.19&47.18&23.49&23.33&19.76\\
C&-100.73&-111.12&-103.53&-101.37&-102.00&16.2&10.82&13.50&13.54&13.39&25.57&51.22&28.15&28.11&24.83\\
N&-129.04&-140.08&-129.30&-127.27&-126.99&20.8&15.57&17.12&16.94&17.15&-&-&-&-&-\\					
O&-192.36&-213.81&-195.95&-191.96&-192.62&59.96&57.58&62.73&62.09&62.43&73.06&75.00&79.11&79.32&73.59\\
F&-256.96&-288.71&-262.19&-258.84&-258.49&63.32&57.15&64.43&63.93&64.42&77.88&78.40&81.92&81.48&77.49\\
Ne&-322.2&-329.69&-329.34&-325.56&-&66.37&66.69&65.80&65.32&-&-&-&-&-&-\\																		
Mg&-34.06&-35.09&-35.00&-35.06&-&-&-&-&-&-&-&-&-&-&-\\											
Al&-58.12&-60.70&-60.31&-60.23&-&17.55&13.22&18.20&17.95&-&14.79&25.40&17.41&18.65&-\\	
Si&-85.46&-90.34&-89.72&-89.22&-&18.37&14.06&18.59&18.42&-&16.9&27.32&17.93&19.31&-\\	
P&-113.07&-122.25&-119.38&-118.24&-&17.83&13.22&17.34&17.07&-&-&-&-&-&-\\						
S&-168.12&-181.11&-179.51&-177.06&-&45.46&42.97&47.21&46.57&-&43.42&44.83&45.82&47.53&-\\	
Cl&-223&-247.70&-237.76&-235.85&-&43.66&27.54&43.58&42.96&-&41.47&36.11&42.57&44.04&-\\
\hline
MAD$_1$&&12.91&3.41&1.83&1.6&&4.34&2.07&1.95&2.38&&14.03&4.24&4.14&0.56\\
MAD$_2$&&9.23&6.64&5.64&-&&6.37&0.64&0.6&-&&6.95&1.79&3.24&-\\	
MAD&&11.21&4.9&3.59&1.6&&5.26&1.42&1.34	&2.38&&10.49&3.02&3.69&0.56\\
		\hline
		\hline
    \end{tabular}
      \label{table:valcorr}
\end{table*}

\begin{table*}
    \centering
    %\small
    \footnotesize  
    \caption{\small{The number of electrons and averaged runtime of a single training iteration for FermiNet with different ECPs for elements from Li to Cl. The average runtime is calculated based on 1000 training iterations. }}
	~\\
	
    \begin{tabular}{ c@{\hspace{0.3cm}} c@{\hspace{0.3cm}} c@{\hspace{0.3cm}} c@{\hspace{0.3cm}} c@{\hspace{0.3cm}} c @{\hspace{0.3cm}}c@{\hspace{0.3cm}} c@{\hspace{0.3cm}} c@{\hspace{0.3cm}} c@{\hspace{0.3cm}} c @{\hspace{0.3cm}}c}
		\hline
		\hline
		&\multicolumn{2}{c}{AE} &\multicolumn{2}{c}{ccECP}&\multicolumn{2}{c}{STU}&\multicolumn{2}{c}{BFD}&\multicolumn{2}{c}{eCEPP}\\ %\cline{2-9}
		\cmidrule(lr){2-3}\cmidrule(lr){4-5}\cmidrule(lr){6-7}\cmidrule(lr){8-9}\cmidrule(lr){10-11}
		Element&Nelec &time(s)&Nelec&time(s)&Nelec&time(s)&Nelec&time(s)&Nelec&time(s)\\		\hline
		Li &3&0.035&1& 0.018&1 &0.020 &1&0.019&1&0.021\\
		Be&4& 0.041&2& 0.034&2 & 0.036& 2&0.037&2&0.036\\  
		B&5&0.051&3&0.043 & 3&0.063 &3&0.044&3&0.053\\
		C&6&0.062&4&0.058 &4 &0.095 &4&0.064&4&0.076\\
		N&7&0.075&5&0.079 &5 &0.137 &5&0.069&5&0.108\\
		O&8&0.088&6&0.104 &6 & 0.188&6&0.109&6&0.147\\
		F&9&0.114&7& 0.133&7 & 0.252&7&0.134&7&0.193\\
		Ne&10&0.131&8&0.168 &8 &0.321 &8&0.169&-&-\\
		Na&11&0.157&1&0.021 &1 &0.022 &1&0.032&-&-\\
		Mg&12&0.181&2& 0.037& 2&0.041 &2&0.047&-&-\\
		Al&13&0.214&3& 0.052& 3& 0.063&3&0.060&-&-\\
		Si&14&0.247&4& 0.076& 4& 0.095&4&0.077&-&-\\
		P&15&0.282&5& 0.108&5 &0.136 &5&0.108&-&-\\
		S&16&0.310&6&0.148 & 6&0.188 &6&0.153&-&-\\
		Cl&17&0.364&7& 0.195& 7&0.252 &7&0.194&-&-\\
		\hline
		\hline
    \end{tabular}
    \label{table:10}
\end{table*}

\begin{table*}[ht]
    \centering
    %\small
    \footnotesize 
\caption{\small Ground state energies for the first two row atoms and CH$_4$. $\Delta E$ represents the differences between the results of the reference FermiNet and the experimental values. 
}

    ~\\
    
\begin{tabular}{ c@{\hspace{0.4cm}} c @{\hspace{0.4cm}} c@{\hspace{0.4cm}}  c@{\hspace{0.4cm}} }
	\hline
		\hline
		&Reference FermiNet$(E_{h})$\cite{ferminet,spencer2020better}&Exact$(E_{h})$\cite{chakravorty1993ground}& $\Delta E_{2}(mE_{h})$\\ 
		\hline
		Li&-7.47798(1)\{20\}&-7.47806&0.08(1) \\
        Be&-14.66733(3)\{20\}&-14.66736 &0.03(3) \\   
		B&-24.6537(3)\{20\}&-24.65391&0.21(30) \\
		C&-37.84471(5) \{20\}&-37.845&0.29(5)  \\
		N&-54.58882(6)\{20\}&-54.5892&0.38(6)  \\
		O&-75.06655(7)\{20\}&-75.0673& 0.75(7) \\
		F&-99.7329(1)\{20\}&-99.7339 & 1.0(1)  \\ 
        Ne&-128.9366(1)\{20\}&-128.9376& 1.0(1) \\
		Na& - &-162.2546& - \\
		Mg& - &-200.053& - \\  
		Al&-&-242.346&  -\\
		Si&- &-289.359 & - \\
		P&-341.2561(1)\{30-40\} &-341.259 &2.9(1)  \\
		S& -398.1049(1)\{30-40\}&-398.11& 5.1(1) \\
		Cl& -460.1452(1)\{30-40\} &-460.148&2.8(1) \\
		CH$_4$&-40.51400(7)\{20\}&-& - \\
		\hline
		MAD&&&1.32\\
		\hline
		\hline
    \end{tabular}
    \label{table:refFerminet}	
\end{table*}
%\newpage
\nocite{*}
\bibliography{manu_clean}% Produces the bibliography via BibTeX.

\end{document}